%% file: example.tex
\documentclass{aa}  
\usepackage[utf8]{inputenc}

\input{fde_emlines}

\usepackage{graphicx}
\usepackage{url}
\usepackage{mathptmx}
\usepackage{amsmath}
\usepackage{anyfontsize}
\usepackage{tensor} 
\usepackage[dvipsnames,hyperref]{xcolor}
\usepackage{multirow}
\usepackage[colorlinks=True, citecolor=blue, linkcolor = cyan, urlcolor=blue]{hyperref} 
\usepackage{natbib}

\hypersetup{
    pdftitle={NGC 424 Marconcini MIRI},
    }
\usepackage[T1]{fontenc}
\newcommand{\msun}{\ensuremath{\mathrm{M_\odot}}\xspace}

\newcommand{\kms}{\ensuremath{\mathrm{km\,s^{-1}}}\xspace}

\newcommand{\MOKA}{\ensuremath{\mathrm{MOKA^{3D}}}}
\def\arcsec{$^{\prime\prime}$}
\def\arcmin{$^{\prime}$}
\let\oldAA\AA
\renewcommand{\AA}{\text{\oldAA}\xspace}
\let\oldarcsec\arcsec
\renewcommand{\arcsec}{\text{\oldarcsec}\xspace}
\newcommand{\hiisi}{{H$_2$}{S(1)}\,}
\newcommand{\hiisii}{{H$_2$}{S(2)}\,}
\newcommand{\hiisiii}{{H$_2$}{S(3)}\,}
\newcommand{\hiisiiii}{{H$_2$}{S(4)}\,}
\newcommand{\hiisiiiii}{{H$_2$}{S(5)}\,}

\begin{document}

   \title{MIRACLE I.: Unveiling the Multi-Phase, Multi-Scale physical properties of the Active Galaxy NGC 424 with MIRI, MUSE, and ALMA}
   \titlerunning{Multi-phase and Multi-scale view of NGC 424}

\author{
C. Marconcini\inst{1,2}\thanks{e-mail: cosimo.marconcini@unifi.it}
\and A. Feltre \inst{2}
\and I. Lamperti \inst{1,2}
\and M. Ceci \inst{1,2}
\and A. Marconi \inst{1, 2}
\and L. Ulivi \inst{1, 2, 3}
\and F. Mannucci \inst{2}
\and G. Cresci \inst{2}
\and F. Belfiore \inst{2}
\and E. Bertola \inst{2}
\and S. Carniani \inst{4}
\and Q. D'Amato \inst{2}
\and J.A. Fernandez-Ontiveros \inst{5,6}
\and J. Fritz \inst{7}
\and M. Ginolfi \inst{1,2}
\and E. Hatziminaoglou \inst{8,9,10}
\and A. Hernán-Caballero \inst{5}
\and M. Hirschmann \inst{11,12}
\and M. Mingozzi \inst{13}
\and A.F. Rojas \inst{14}
\and G. Sabatini\inst{2}
\and F. Salvestrini \inst{12, 15}
\and M. Scialpi \inst{1, 2, 3}
\and G. Tozzi \inst{16}
\and G. Venturi \inst{4,2}
\and A. Vidal-García \inst{17}
\and C. Vignali \inst{18,19}
\and M.V. Zanchettin \inst{2}
\and A. Amiri \inst{20}
}
\institute{
Dipartimento di Fisica e Astronomia, Università degli Studi di Firenze, Via G. Sansone 1,I-50019, Sesto Fiorentino, Firenze, Italy
\and
INAF - Osservatorio Astrofisico di Arcetri, Largo E. Fermi 5, I-50125, Firenze, Italy
\and
University of Trento, Via Sommarive 14, Trento, I-38123, Italy
\and
Scuola Normale Superiore, Piazza dei Cavalieri 7, Pisa, I-56126, Italy
\and
Centro de Estudios de Física del Cosmos de Aragón (CEFCA), Plaza San Juan 1, 44001, Teruel, Spain
\and
Istituto di Astrofisica e Planetologia Spaziali (INAF-IAPS), Via Fosso del Cavaliere 100, 00133, Roma, Italy
\and
Instituto de Radioastronomía y Astrofísica, UNAM, Campus Morelia, A.P. 3-72, C.P. 58089, Mexico
\and
ESO, Karl-Schwarzschild-Str 2, D-85748 Garching bei München, Germany
\and
Instituto de Astrof\'{i}sica de Canarias, 38205 La Laguna, Tenerife, Spain
\and
Departamento de Astrof\'{i}sica, Universidad de La Laguna, 38206 La Laguna, Tenerife, Spain
\and
Institute for Physics, Laboratory for Galaxy Evolution and Spectral Modelling, Ecole Polytechnique Federale de Lausanne, Observatoire de Sauverny, Chemin Pegasi 51, CH-1290 Versoix, Switzerland
\and 
INAF, Osservatorio Astronomico di Trieste, Via Tiepolo 11, I-34131 Trieste, Italy 
\and 
AURA for ESA, Space Telescope Science Institute, 3700 San Martin Drive, Baltimore, MD 21218, USA
\and
Departamento de Física, Universidad Técnica Federico Santa María, Vicu\~{n}a Mackenna 3939, San Joaqu\'in, Santiago de Chile, Chile
\and
IFPU - Institute for Fundamental Physics of the Universe, via Beirut 2, I-34151 Trieste, Italy
\and
Max-Planck-Institut für extraterrestrische Physik (MPE), Gie\ss{}enbachstraße 1, 85748 Garching, Germany
\and
Observatorio Astronómico Nacional, C/ Alfonso XII 3, E-28014 Madrid, Spain
\and
Dipartimento di Fisica e Astronomia, Alma Mater Studiorum, Università degli Studi di Bologna, Via Gobetti 93/2, 40129 Bologna, Italy
\and
INAF–Osservatorio di Astrofisica e Scienza dello Spazio di Bologna, Via Gobetti 93/3, 40129 Bologna, Italy
\and
Department of Physics, University of Arkansas, 226 Physics Building, 825 West Dickson Street, Fayetteville, AR 72701, USA
}

   \date{Received~\today; Accepted~NNN~NN, NNNN}

   \abstract
   {We present the analysis of the multi-phase gas properties in the Seyfert II galaxy NGC 424, using spatially resolved spectroscopic data from JWST/MIRI, part of the Mid-InfraRed Activity of Circumnuclear Line Emission (MIRACLE) program, as well as VLT/MUSE and ALMA. We trace the properties of the multi-phase medium,  from cold and warm molecular gas to hot ionized gas, using emission lines such as CO~(2-1), \hiisi, \OIIIL, [Ne \text{III}]15.55$\mu$m, and [Ne \text{V}]14.32$\mu$m. These lines reveal the intricate interplay between the different gas phases within the circumnuclear region, spanning approximately 1.4 $\times$ 1.4 kpc$^2$, with a resolution of 10 pc. Exploiting the multi-wavelength and multi-scale observations of gas emission we model the galaxy disc rotation curve from scales of a few parsec up to $\sim$ 5 kpc from the nucleus and infer a dynamical mass of M$_{\rm dyn}$ = (1.09 $\pm$ 0.08) $\times$ 10$^{10}$ \msun with a disc scale radius of R$_{\rm D}$ = (0.48 $\pm$ 0.02) kpc.
   We detect a compact ionised outflow with velocities up to 10$^{3}$~\kms, traced by the \OIII, \NeIIImu, and \NeVmu transitions, with no evidence of cold or warm molecular outflows. We suggest that the ionised outflow might be able to inject a significant amount of energy into the circumnuclear region, potentially hindering the formation of a molecular wind, as the molecular gas is observed to be denser and less diffuse.
   The combined multi-band observations also reveal, in all gas phases, a strong enhancement of the gas velocity dispersion directed along the galaxy minor axis, perpendicular to the high-velocity ionised outflow, and extending up to 1 kpc from the nucleus. Our findings suggest that the outflow might play a key role in such enhancement by injecting energy into the host disc and perturbing the ambient material.}
   
   \maketitle
\section{Introduction}\label{sec.introduction}
Active Galactic Nuclei (AGN) are widely recognized as fundamental drivers of galaxy evolution through the AGN feedback process. The energy released by gas accretion onto supermassive black holes (SMBHs) can generate powerful winds and outflows that propagate through the host galaxy, influencing the surrounding interstellar medium (ISM). AGN-driven outflows are now considered a key mechanism for regulating star formation and shaping the overall mass distribution in galaxies \citep[e.g.][]{Fabian2012, Kormendy2013, King_pounds2015, Cicone2018, Harrison2018}

AGN-driven outflows are inherently multi-phase, comprising gas in the ionised, atomic, and molecular states, each contributing differently to feedback processes. Ionised outflows, often traced by optical emission lines like \OIIIL, can reach velocities of several hundred to a few thousand km s$^{-1}$, transporting warm (T $\sim$ 10$^{4}$ K), low-density gas (N${\rm e}$ $\sim$ 10$^{2-4}$ cm$^{-3}$) out of the nucleus and up to kpc scales \citep[e.g.][]{ Carniani2015, Bae2016, Woo2016, Fiore2017, Cicone2018, Venturi2018_magnum, Harrison2018, Venturi2021_turmoil, Cresci2023}. These outflows are critical in dispersing gas from the nuclear regions, though their efficiency in quenching star formation is still debated both in observations and simulations \citep[e.g. see][]{Cresci2015, Balmaverde2016, Zubovas2017, Combes2017, Mulcahey2022, Piotrowska2022, Belli2024}. Molecular outflows, on the other hand, represent colder, denser phases of gas and are frequently observed via CO transitions. They are typically slower, with velocities on the order of a few hundred km s$^{-1}$, but can carry a substantial amount of mass (M$_{\rm out}$ $\sim$ 10$^{7}$-10$^{9}$ \msun) and are possibly more directly involved in regulating star formation \citep[e.g.][]{Feruglio2010, Fiore2017, Carniani2015, GarciaBurillo2019, Fluetsch2019}.

Understanding the energetics of these outflows — specifically their mass outflow rates, momentum, and energy flux — is essential for constraining their impact on galactic scales. Recent studies have shown that the coupling efficiency between the SMBH and the surrounding gas can vary depending on the phase of the outflow. Indeed, although the cold molecular phase carries the bulk of the outflowing gas mass, the kinetic energies are higher in the ionised phase \citep{Rupke2017, Vayner2021, Riffel2023, harrison2024}. Multiphase outflows are thus a key feature of AGN feedback, as they link the energetic output of the AGN to the ambient ISM across a wide range of temperatures and densities \citep[e.g.][]{Rupke2013, Cazzoli2018}.


\begin{figure*}
	\includegraphics[width=\linewidth]{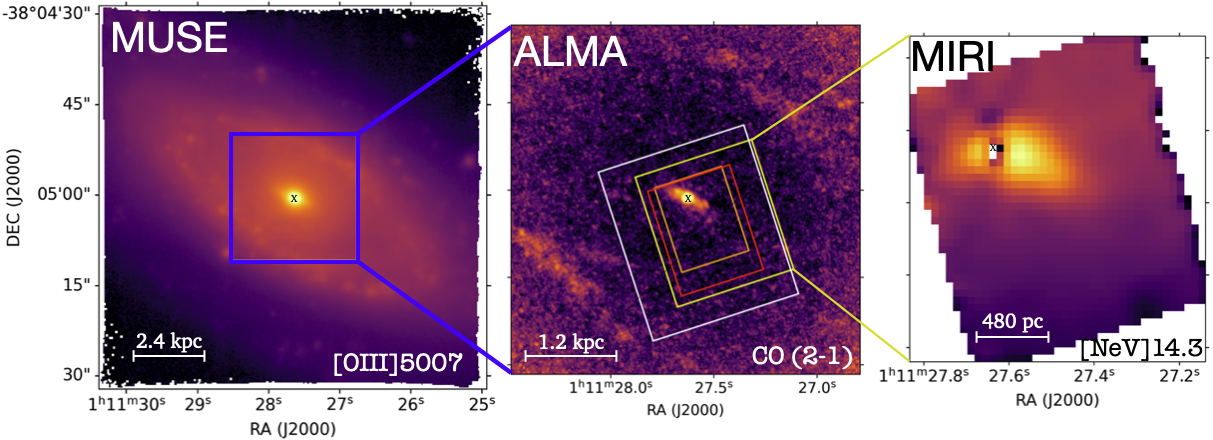}
    \caption{NGC 424 emission line images from MUSE WFM (left), ALMA (middle) and point-source subtracted MIRI MRS Ch3 medium (right). The images are obtained integrating the total \OIIIL, CO~(2-1), and \NeVmu emission lines, respectively. Blue square in MUSE image represents the ALMA FOV considered in this work. Orange, red, yellow, and white rectangles in ALMA image represent the FOVs of MIRI MRS Channels 1, 2, 3, and 4, respectively. Black crosses mark the nucleus position.}
    \label{fig:FOV_telescopes}
\end{figure*}

Recent high-resolution observations from facilities such as the Atacama Large Millimeter Array \citep[ALMA;][]{Wootten2009} have investigated the role of cold molecular outflows in AGN, showing that this gas phase can be accelerated to high velocities and can extend across several kpc \citep{Fluetsch2019}. Additionally, ionised outflows are a ubiquitous manifestation of AGN feedback, mainly traced through optical emission lines such as \OIIIL and \Halpha \citep[e.g.][]{Harrison2014, Fiore2017}. Finally, warm molecular outflows (T = 10–10$^{2}$ K, n$_{\rm e}$ $\geq$ 10$^{3}$ cm$^{-3}$) are typically traced by rotational transitions of H$_2$ in the Near Infrared (NIR) using the  Spectrograph for INtegral Field Observations in the Near Infrared (SINFONI; \citet{Eisenhauer2003}), and in the Mid-Infrared using \textit{Spitzer} and more recently by JWST \citep{Wright2023} Mid-Infrared observations. H$_2$ emission is both observed and predicted to be particularly rich in shocked gas and therefore representing a valuable tracer of warm molecular outflowing gas \citep[][]{Hill2014, Richings2018, Richings2018b, Riffel2020}. Thus, the role of multiphase outflows spans from the circumnuclear scale up to galactic scales, since they can sweep away the ambient material and also affect the conditions of the host halo, potentially halting cooling flows and limiting the accretion of new gas onto the galaxy.

In addition to their feedback potential, multiphase outflows provide important clues about the physics of gas accretion and ejection around SMBHs. The co-existence of gas in very different physical conditions, ranging from ionised, high-temperature plasma to cold molecular clouds indicates that AGN-driven winds may entrain gas from different regions of the ISM or cool as they propagate outward \citep{Zubovas2024}. This entrainment process allows molecular gas to survive in extreme environments, where heating from shocks and radiation fields might otherwise be expected to dissociate or ionize the gas \citep{Richings2018, Chen2024}. Studying these outflows in detail, and in particular with high spatial resolution, is therefore crucial for understanding how energy is transferred across the ISM and how AGN feedback can shape the evolution of galaxies \citep{Ward2024, Sivasankaran2024, Byrne2024}.

Recent JWST/MIRI \citep{Rieke2015b} observations already started to revolutionize our comprehension of the Mid-IR activity of local AGN by allowing Integral Field Spectroscopy (IFS) in this spectral region \citep{garciabernete2022, Armus2023, Bajaj2024}. Indeed, the Medium resolution Spectrometer (MRS, \citealt{Labiano2021}) exceptional sensitivity and spatial resolution allow to detect high-ionization transitions such as [Ne \text{V}]14.3 $\mu$m, [Ne \text{V}]24.3 $\mu$m and \OIVmu, which is proving to be crucial to probe the AGN activity in the circumnuclear region \citep{Lai2022, Zhang2024, Riffel2024} thanks to the nearly unbiased view through dust attenuation \citep{Gordon2023, garciabernete2024, Donnan2024}. This breakthrough capability is pivotal for unveiling the intricate interplay between the AGN radiation field and the circumnuclear ambient medium. Moreover, combining the Mid-Infrared Instrument (MIR) MRS with the optical Multi Unit Spectroscopic Explorer (MUSE) at the ESO Very Large Telescope \citep[VLT;][]{Bacon2010} and the sensitivity and spatial resolution in the mm-regime of ALMA, pave the way for a complete characterization of the multi-phase properties of galactic outflows. In particular, the synergy among these facilities enables to simultaneously trace the ionized, warm, and cold gas components, offering a holistic view of the feedback processes regulating the star formation and driving the galaxy evolution. Such a multi-wavelength and multi-scale approach marks a significant step forward in our ability to disentangle the complex mechanisms governing AGN-driven outflows and their impact on the host galaxy.

In this paper, we present the first data of our Mid-InfraRed Activity of Cicumnuclear Line Emission (MIRACLE; PID: 6138, Co-PIs: C. Marconcini and A. Feltre) program, aimed at observing a sample of seven nearby AGN in the 5--28 $\mu$m wavelength range with the MIRI MRS onboard the JWST telescope. The goal of this project is to characterize the interplay between the AGN and the host galaxy in the circumnuclear region and across a wide range of wavelength and spatial scales.

This paper focuses on a comprehensive analysis of the multiphase gas properties in the first observed target of our program, i.e. NGC 424, using a combination of infrared, optical and millimeter data from MIRI on the JWST, MUSE, and ALMA, respectively. These observations allow the multi-phase gas to be traced across a wide range of temperatures and densities, providing a detailed view of its kinematics, structure, and energetics.
\input{NGC424_commissioning}

NGC 424 is a nearby heavily obscured Seyfert-II galaxy (z = 0.01175 $\pm$ 0.00006\footnote{We estimated the redshift fitting the \Hbeta and \OIIIall emission lines from the integrated spectrum of MUSE data extracted from a circular region of 1\arcsec centred on the nucleus.}, D = 51 Mpc, 1\arcsec $\sim$ 230 pc), poorly studied in any gas phase. Nevertheless, the proximity of NGC 424 makes it an ideal target for high-resolution, multi-wavelength studies aimed at tracing the full extent of the gas properties across a wide  wavelength range to understand the interplay between its different gas phases. In the X-rays, \citet{Ricci2017} estimated an intrinsic 2-10 keV X-ray luminosity of log(L$_{\rm X}$/erg s$^{-1}$) =  43.77 and a column density of log(NH/cm$^{-2}$) = 24.33 $\pm$ 0.01. Additionally, \citet{Kakkad2022} used archival MUSE observations to estimate the ionised outflow energetic traced by the \OIIIL emission line. They found an average outflow velocity of 814 $\pm$ 70 \kms and a mass outflow rate of 26.3 $\pm$ 1.3 \msun yr$^{-1}$. In this paper we present the first observations of our MIRACLE program, providing a comprehensive energetic analysis of the multi-phase outflow across various scales in NGC 424. 

This paper is organized as follows. In Section 2, we describe the observations and data reduction procedures for MIRI, MUSE, and ALMA. In Section 3 we present the data analysis for each instrument, discussing the spectroscopic analysis of specific emission lines tracing different gas phases. In Section 4 we discuss our results and present a detailed morphological and kinematic analysis of the multi-phase gas properties across different scales. In Section 5, we discuss the implications of our findings for AGN feedback and the interplay among different gas phases. Finally, in Section 6 we summarize our findings.

\section{Observations and data reduction}\label{sec.obs_and_data_red}
We combine new observations of NGC 424 obtained with the MRS as part of the MIRACLE survey (Marconcini et al. in prep.) with archival observations from VLT/MUSE and ALMA. The Field of View (FOV) covered by each instrument is shown in Fig. \ref{fig:FOV_telescopes}. The observations and data reduction pipeline adopted for each telescope are discussed in the following sections.

\subsection{JWST/MIRI}\label{sec_miri_reduction}
The JWST/MIRI observations of NGC 424 are part of the GO Cycle 3 MIRACLE program 6138. The data were acquired with MIRI \citep[][]{Rieke2015, Rieke2015b, Labiano2021} in MRS mode \citep[][]{Wells2015} on 2024 October 20 UT (see Tab. \ref{tab:MIRI_commissioning_NGC424} for a summary of the observations). We used the MRS with all channels 1-4, covering the spectral range 4.9-28.1 $\mu$m, with three grating settings, SHORT (A), MEDIUM (B), and LONG (C) for each channel. For each sub-channel, the science exposure time was 555 s and a four-pointing dither pattern was used to improve the pixel sampling. We opted for the FASTR1 readout pattern to optimize the dynamic range expected in the observations. Because our source is extended, we linked the science observation to a dedicated background with the same observational parameters in all three grating settings. We downloaded the uncalibrated science and background observations through the Barbara A. Mikulski Archive for Space Telescopes (MAST) portal and the data reduction process was performed using the JWST Science Calibration Pipeline \citep{Bushouse2022_pipeline} version $1.16.0$. We applied all the three stages of the pipeline processing, which include  \texttt{CALWEBB\_DETECTOR1},  \texttt{CALWEBB\_SPEC2}, and \texttt{CALWEBB\_SPEC3} \citep[see][]{Morrison2023,Patapis2024}. 

\begin{figure*}
	\includegraphics[width=\linewidth]{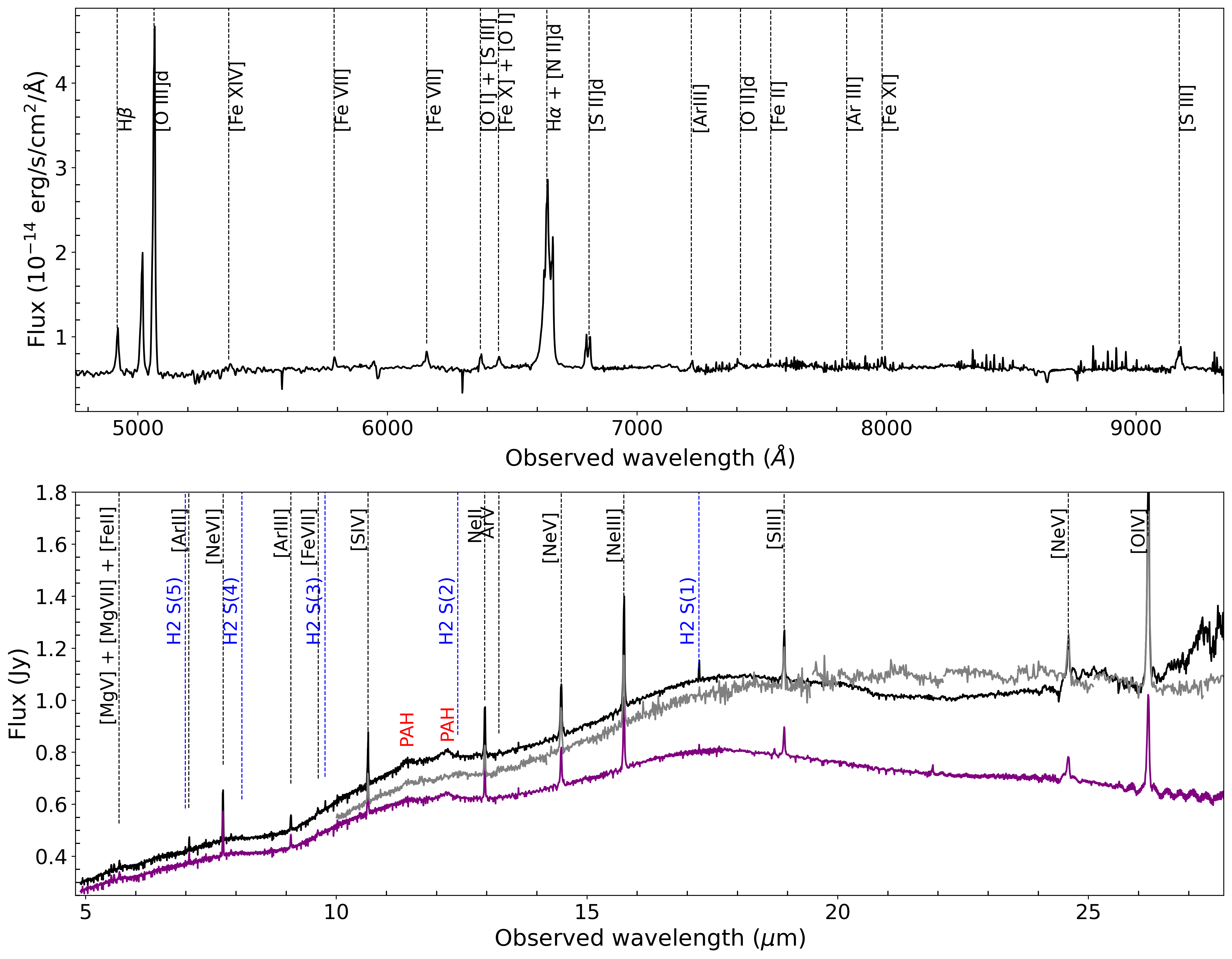}
    \caption{MUSE (top panel) and MIRI MRS (bottom panel) integrated spectra for NGC 424 extracted from the same region, i.e. the MIRI MRS Ch 1 (orange square in Fig. \ref{fig:FOV_telescopes}). All the emission lines detected in each spectrum are highlighted. Warm molecular hydrogen transitions in the MIRI MRS spectrum are shown in blue. Black and purple spectra in the bottom panel represent the integrated MRS spectrum extracted from the entire FOV and from the 1 arcsec$^2$ nuclear region of NGC 424, respectively. Red labels mark the positions of PAHs features at 11.25 and 12 $\mu$m. Grey spectrum in the bottom panel is the \textit{Spitzer} IRS high-resolution extracted in full mode normalised to the MIRI MRS integrated spectrum.}
    \label{fig:spectra_MIRI}
\end{figure*}

Additional fringe corrections were made during stages 2 and 3, using the standard pipeline code. The \texttt{CALWEBB\_DETECTOR1} pipeline corrects the raw detector ramps of each exposure for multiple electronic artifacts such as bad pixels and cosmic-rays, producing slope images with the count rate in Data Number (DN) per second in each pixel. The resulting slope images are then processed with the \texttt{CALWEBB\_DETECTOR2} step, where each image is corrected for distortion and then calibrated both in wavelength and flux. We refer the reader to \citet{Argyriou2020, Argyriou2023} for a detailed description of all the detector-level steps performed during this stage. A first residual fringe correction is applied during this stage. The results of stage 2 are calibrated slope images in units of MJy sr$^{-1}$. Then, each calibrated image goes through the final \texttt{CALWEBB\_SPEC3} stage which performs background subtraction to each single-exposure observation. In this work, we subtracted the background emission from the 2D science images using a spaxel-by-spaxel background frame generated from our dedicated background observations. Then, the main step of the third stage is to create the final data cubes by combining the flux-calibrated, dithered science images in a composite 3D data cube. To assemble the final data cubes, we used the exponential modified-Shepard method (EMSM) weighting function, which has proven to be more efficient in reducing the drizzling effect \citep{Law2023_drizzle} and thus improve the final output. We combined all the final data cubes on the detector plane using the \textit{skyalign} orientation provided by the JWST pipeline, i.e. cubes are oriented according to the world coordinates RA, DEC. Based on these steps, the results of the pipeline process are 12 data-cubes, one per each sub-band, which span progressively larger FOV, from 3.2\arcsec $\times$ 3.7\arcsec\ in Channel 1 to 6.6\arcsec $\times$ 7.7\arcsec\ in Channel 4. Moreover, each sub-Channel data cube has a different spaxel sampling, i.e. 0.13, 0.17, 0.2, and 0.35 \arcsec/pxl, from Channel 1 to Channel 4 respectively. The MIRI integrated spectrum is obtained combining the emission from the 12 reduced data-cubes, taking into account the different pixel sizes and ensuring a proper flux conservation among different bands applying a scaling factor to align the flux levels between adjacent bands and stitch the spectra. A comprehensive description of the procedure for stitching together spectra covering different bands and accounting for variable pixel sizes and FOV is presented in Ceci et al. (in prep.). Finally, to investigate the spatially resolved extended emission in each band we performed a detailed Point Spread Function (PSF) subtraction procedure to each data-cube, using the \textit{WebbPSF} tool (see Appendix \ref{sec_app_PSF_miri_sub} for more details).

\subsection{MUSE}\label{sub_sec_muse_observations}
NGC 424 IFS archival data obtained with MUSE (ID: 095.B-0934, P.I. S. Juneau) are part of the extended Measuring AGN Under Muse (MAGNUM) sample \citep{Cresci2015, Venturi2017, Venturi2018_magnum, Mingozzi2019, Marconcini2023}. The sample was selected by cross-matching the optically selected AGN samples of \citet{Maiolino1995} and \citet{Risaliti1999}, and Swift-BAT 70-month Hard Xray Survey \citep{Baumgartner2013}, choosing only sources observable from Paranal Observatory (70$^{\circ}$ $\leq$ $\delta$ $\leq$ 20$^{\circ}$) and with a luminosity distance D $\leq$ 50 Mpc. We retrieved the data from ESO archive\footnote{\url{https://archive.eso.org/cms.html}}, already reduced with the standard MUSE pipeline (v1.6) and with an average PSF FWHM of 0.8\arcsec. The final data-cube consists of 322 $\times$ 318 spaxels, with a spatial sampling of 0.2\arcsec pixel$^{-1}$, covering the spectral range between 4750 to 9350 \AA, thus tracing the rest-frame optical wavelength range. The MUSE spectral resolution spans from 1750 at 4650 \AA to 3750 at 9300 \AA. The observations field of view of 1\arcmin $\times$ 1\arcmin\ corresponds to a region of 14 kpc $\times$ 14 kpc centred in the nucleus of NGC 424.

\subsection{ALMA}\label{sub_sec_alma_observations}
To investigate the cold molecular gas phase in NGC 424 we analysed archival ALMA 12-m array band 6 observations from program 2021.1.01150.S (P.I. A. Rojas) covering the observed spectral window of [226.91, 228.79] GHz which allowed us to trace the CO~(2-1) emission at 230.538 GHz rest-frame. For the ALMA data considered in this work, the FOV is 38\arcsec in diameter, with the largest recoverable scale of 11.4\arcsec.
We requested the calibrated measurement sets from the European ALMA Regional Centre \citep[ARC;][]{Hatziminaoglou2015}. To reduce and analyse the data we used the Common Astronomy Software Applications (CASA) package version v6.1.1 \citep[][]{McMullin2007, CASATeam2022}. This program includes observations with two configurations, with beam sizes of 0.2\arcsec and 1\arcsec, respectively. We combined the two measurement sets to obtain the best compromise in terms of spatial resolution and sensitivity. In particular, for the CO~(2–1) spectral window, we subtracted a constant continuum level estimated from the emission line free channels in the \textit{uv} plane\footnote{The continuum was subtracted using the line free spectral channels in the spectral windows [226.91, 228.79] GHz and [229.29, 231.27] GHz}. The data were cleaned using the CASA task \textsc{tclean} CASA task using a \textsc{Briggs} weighting scheme with \textsc{robust} = 0.5, to achieve a $\sim$110 pc resolution (beam size FWHM 0.48\arcsec $\times$ 0.44\arcsec, beam PA = 1$^{\circ}$). The final reduced data cube has a spectral channel width of $\sim$ 5 km s$^{-1}$, a pixel size of 90 mas, and an rms of 0.94 mJy/beam per channel.


\section{Data analysis}\label{sec.data_analysis}
In this section we outline the general spectroscopic routine used to perform the emission line analysis of the MUSE, ALMA, and MIRI data-cubes. Overall, we analysed the MIRI and MUSE data-cubes using a set of tailored python scripts in order to subtract the continuum and fit the emission lines with multiple Gaussian components where needed. Fig. \ref{fig:spectra_MIRI} shows the NGC 424 MUSE (top panel) and MIRI (bottom panel) integrated spectra with the detected optical and Mid-IR emission lines, extracted from the smallest MIRI/MRS FOV (i.e. Channel 1). The integrated flux of each detected emission line in the MIRI/MRS spectral coverage, together with the maximum number of Gaussian components used to reproduce each line profile, are listed in Tab. \ref{tab:line_fluxes_miri}. For comparison, in bottom panel in Fig. \ref{fig:spectra_MIRI} we show the \textit{Spitzer} \citep{Werner2004} Infrared Spectrometer \citep[IRS;][]{Houck2004} high-resolution spectrum of NGC 424 extracted in full mode (PID: 30291; P.I.: G.Fazio) and retrieved from the Combined Atlas of Sources with \textit{Spitzer} IRS \citep[CASSIS][]{Lebouteiller2011, Lebouteiller2015}. The \textit{Spitzer}/IRS spectrum reveals the main mid-IR ionised transitions as well as the \hiisi transition, together with the source Mid-IR continuum. As shown in bottom panel in Fig. \ref{fig:spectra_MIRI} we find evidence of polycyclic aromatic hydrocarbon (PAH) features at 11.25 and 12 $\mu$m, with higher Equivalent Width (EW) with respect to the \textit{Spitzer} spectrum. In particular, we performed a local continuum fit, integrated the PAH flux, and estimated EW$_{11 \mu m}$ = 0.007 $\mu m$, and EW$_{12 \mu m}$ = 0.01 $\mu m$, for the PAH features at 11.25 and 12 $\mu$m, respectively. Interestingly, \citet{Wu2009} and \citet{Tommasin2010} reported a weak detection (EW = 0.01 and 0.001 $\mu$m, respectively) of the PAH feature at 11.25 $\mu$m from the same \textit{Spitzer} data shown in Fig. \ref{fig:spectra_MIRI} \citep[see also][]{Hernan_Caballero2011}. Nevertheless, they found no evidence of the PAH feature at 12 $\mu$m. A comprehensive analysis of PAH properties in the entire MIRACLE sample will be presented in a dedicated paper. The MIRI/MRS spectrum also reveals no silicate absorption or emission features at 9.7 or 18 $\mu$m. Silicate features are sensitive to the geometry of the dust distribution along the line of sight and to the amount of dust obscuration \citep{Imanishi2003, Imanishi2007, Hernan_Caballero2011}. Therefore, in highly-obscured AGN, the silicate absorption is expected due to the dense, dusty torus \citep[][]{Spoon2007}. However, its non-detection in NGC 424 suggests a clumpy dust distribution in the nuclear region which allows the Mid-IR radiation to escape with minimal absorption \citep{Nenkova2008a, Nenkova2008b}. Additionally, the absence of silicate emission, which is scarcely observed in Type- II AGN due to the obscured hot torus surface, supports the idea that the AGN-heated dust is hidden from our line of sight \citep[for another perspective see][]{Hatziminaoglou2015b}.

\input{line_fluxes}


\subsection{JWST/MIRI MRS emission line fitting}\label{sec.em_line_analysis_MIRI}
To extract the highly-ionised and warm molecular gas properties from MRS data we performed a preliminary Voronoi tessellation \citep{Cappellari2003} in order to achieve an average signal-to-noise (S/N) ratio of 20 per wavelength channel on the continuum around emission lines. In this work we focus on three emission lines tracing the ionised and warm molecular phases. In particular, we considered the [Ne \text{V}]14.3 $\mu$m (ionization potential (IP) of 97 eV), [Ne \text{III}]15.5$\mu$m (IP = 41 eV), and \hiisi 17$\mu$m (hereafter \NeVmu, \NeIII, and \hiisi, respectively) emission lines from Channel 3 (Ch3, hereafter) MEDIUM and LONG, respectively. The choice of these specific emission lines is justified as they are among the strongest emission lines for each category (i.e. high- and low-ionisation potential transitions and molecular H$_2$), and have many further advantages. First, these transitions occur in the same channel and thus are  characterised by the same pixel scale (i.e. 0.2\arcsec/pixel) and similar spatial resolution (see App. \ref{sec_app_PSF_miri_sub}). Second, since the attenuation curve at these wavelengths is almost flat we expect a similar and mild attenuation of the line fluxes \citep[][]{Lai2024, Donnan2024}, thus making the flux-derived properties reliable.  We performed a local continuum subtraction, focusing on two independent spectral regions encompassing a velocity range of $\pm$ 2500 \kms around each emission line \citep[for a similar approach to MIRI MRS data see][]{Bajaj2024}. In particular, we used the Penalized Pixel-Fitting \citep[pPXF,][]{Cappellari2004} code on the previously binned spaxels to fit a first degree polynomial to the continuum underlying the emission line. Moreover, we simultaneously fit the emission line in the selected wavelength range, with one or two Gaussian components, to properly fit the continuum shape. Then, we subtracted the best-fit continuum and obtained a continuum-subtracted model cube which we spatially smoothed with a Gaussian kernel with $\sigma$ = 1 spaxel (i.e. 0.2\arcsec). The spatial smoothing does not affect the MRS spatial resolution at this wavelength, since in Ch3 the pixel size is smaller than the PSF (see also App. \ref{sec_app_PSF_miri_sub}). 

We then analysed the smoothed continuum-subtracted data cube to perform a detailed emission line fitting. In particular, we independently fitted the \NeVmu, \NeIIImu, and \hiisi emission lines with a different number of Gaussians. Due to line asymmetries of the \NeVmu, \NeIIImu ionised transitions, two Gaussians were necessary to reproduce the line profile. On the other hand, the \hiisi was well reproduced using a single Gaussian component. We used the MPFIT fitting tool \citep{Markwardt2009} to fit the emission lines in each spaxel. To decide the optimal minimum number of Gaussian components necessary to reproduce the line profile we applied a reduced $\chi^2$ selection and a Kolmogorov-Smirnoff test on the residuals of to the best fits in each spaxel, using a fiducial \textit{p-value} of 0.8 \citep[for more details on the fitting algorithm see][]{Marasco2020}. 

\begin{figure*}
	\includegraphics[width=\linewidth]{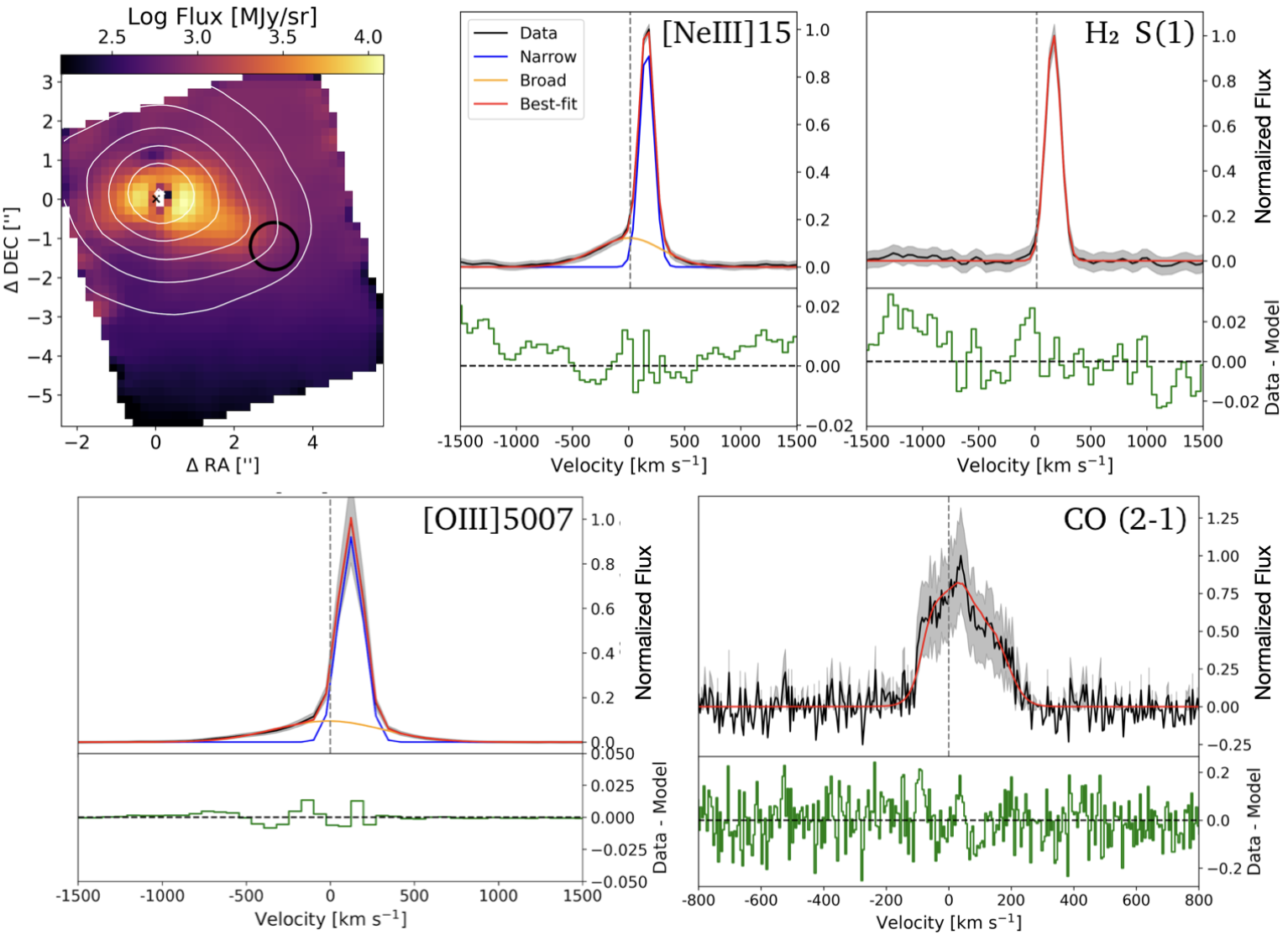}
    \caption{Multi-phase Gaussian fit of the \NeIIImu, \hiisi, \OIIIL, and CO~(2-1) emission lines from MIRI/MRS, MUSE, and ALMA data cubes, respectively. Top-left panel: \NeIIImu emission tracing the red-shifted ionised gas, obtained collapsing the flux in the spectral window 15.56 - 15.57 $\mu$m rest-frame from the PSF subtracted data cube (see App. \ref{sec_app_PSF_miri_sub}). White contours represent arbitrary \OIII flux levels from MUSE data. Top-right panels: Integrated \NeIIImu and \hiisi emission extracted from the black circular aperture of radius 0.6 \arcsec, shown on the left panel. Bottom left: Integrated \OIIIL emission and best-fit extracted from the same aperture shown in the top-left panel. Bottom right: Integrated CO~(2-1) emission extracted from spaxels with S/N $\geq$ 3, covering the nuclear region of NGC 424, as shown in bottom panel in Fig. \ref{fig:disc_emission_4_tracers}. Data and total best-fit model are in black and red, respectively. The sum of the single Gaussian best-fit components is shown in blue (narrow), and orange (broad). Bottom panels show the residuals in green. Dashed vertical gray lines mark the rest-frame wavelength of the transitions. Data and best-fit are normalized to the peak of the observed spectrum.}
    \label{fig:fit_NeIII_H2S1_multigaussian_fit}
\end{figure*}

As a result of the multi-Gaussian fit we obtained a dedicated model cube for each Gaussian component and one for their total best-fit profile. In particular, we considered the Gaussian component with lower width to be representative of the systemic disc kinematics and the broader component to trace the outflowing gas. Fig. \ref{fig:fit_NeIII_H2S1_multigaussian_fit} shows the integrated spectra extracted from a circular aperture with radius of 0.6\arcsec located in the red-shifted side of the galactic disc. Top panels in Fig. \ref{fig:fit_NeIII_H2S1_multigaussian_fit} show the multi-Gaussian best-fit model used to reproduce the \NeIIImu and \hiisi line profiles. We observe that the narrower component of the \NeIIImu line profile follows similar kinematics as the \hiisi\ line. On the other hand, a broader blue-shifted component of the \NeIIImu is detected at high-S/N over the entire Ch3 FOV. The \NeVmu shows a line profile similar to that of \NeIIImu but with a more prominent broader component, which we associate to an AGN-driven wind, possibly due to its higher ionization potential (see Sect. \ref{sec.highly_ionised_outflow} for a detailed discussion). Being more sensitive to the intense AGN radiation field in the circumnuclear region, the \NeVmu transition is well suited to trace the properties of such hypothetical wind.

The emission line fitting analysis was performed on the PSF-subtracted Ch3 data cube due to the outshining emission from the unresolved nucleus. As a consequence, to investigate the spatial features of both the narrow and broad components of emission lines we subtracted the model PSF in each spectra channel following the routine described in App. \ref{sec_app_PSF_miri_sub}.

\subsection{MUSE emission line fitting}\label{sec.em_line_analysis_MUSE}
The routine used to perform the spatially resolved ionised gas analysis from the MUSE data-cube is the same used for the MRS data, discussed in Sect. \ref{sec.em_line_analysis_MIRI}, and tailored to MUSE \citep[for an application of this routine to MUSE data see][]{Venturi2018_magnum, Mingozzi2019, Marasco2020, Marconcini2023, Ulivi2024_sigmaperp}. In particular, we performed the Voronoi tessellation requiring an average S/N per wavelength channel of 100 on the continuum between 5150 \AA and 8900 \AA, masking the emission lines within this spectral window. 
We performed the continuum fitting spanning approximately the entire range of the MUSE spectral coverage, i.e. 4750-9000 \AA, which is optimal as it includes many stellar absorption features.  The stellar continuum fit is performed with the pPXF tool, using a linear combination of synthetic single stellar population (SSP) templates from \citet{Vazdekis2010}. The templates are convolved with the MUSE spectral resolution, then shifted, broadened and combined with a first degree additive polynomial to reproduce the observed features. To account for possible absorptions underlying Balmer emission lines we fitted the SSP templates together with the main gas emission lines, i.e. \Halpha, \Hbeta, \OIIIall, \NIIall, \SIIall. We then subtracted the best-fit continuum in each bin and obtained a continuum-subtracted model cube. 

Similarly to the method described in Sect. \ref{sec.em_line_analysis_MIRI}, we fitted the mentioned emission lines in the continuum-subtracted cube with up to two Gaussian components, tying the velocity and velocity dispersion of each component. Moreover, we fixed the flux ratio between the two lines in each doublet, i.e. \OIIIall, and \NIIall, as given by the Einstein coefficients of the two transitions \citep{Osterbrock2006}. The optimal number of Gaussian components is decided based on the $\chi^2$ minimization and a Kolmogorov-Smirnoff test, as already described in Sect. \ref{sec.em_line_analysis_MIRI}. As a result, we obtain an emission-line model cube for all the mentioned transitions. The bottom left panel in Fig. \ref{fig:fit_NeIII_H2S1_multigaussian_fit} shows the integrated spectrum and best-fit of the \OIIIL emission line extracted from the circular aperture shown in Fig. \ref{fig:fit_NeIII_H2S1_multigaussian_fit}. In the following, we will focus on the brightest (and highest ionization, in the optical) emission line, i.e. the \OIIIL transition (hereafter \OIII) to trace the ionised gas features from the MUSE data.


\subsection{ALMA emission line fitting}\label{sec.em_line_analysis_ALMA}
To analyse the properties of the cold molecular gas in NGC 424 we exploited the CO~(2-1) transition observed in the ALMA band 6 data. Since the continuum was already subtracted in the uv-plane during the data reduction we perform a single-Gaussian fit to the line profile. As for the \hiisi transition, a single Gaussian was sufficient to reproduce the line profile in each spaxel based on the $\chi^2$ and the Kolmogorov-Smirnoff test. Bottom right panel in Fig. \ref{fig:fit_NeIII_H2S1_multigaussian_fit} shows the integrated CO~(2-1) spectrum and best-fit extracted from all spaxels with S/N $\geq$ 3.

\section{Results}\label{sec.results}
In this section, we trace the ionised and molecular gas properties exploiting multi-band data from MIRI, MUSE and ALMA (see Sect. \ref{sec.obs_and_data_red}). We explored the molecular and ionised gas content within NGC 424, from the circumnuclear scale up to $\sim$ 5 kpc, providing a comprehensive analysis of the gas properties across different phases. In particular, we used the \hiisi and CO~(2-1) transitions to study the molecular gas component of the disc, from pc scales up to $\sim$ 1 kpc (see Sects. \ref{sec.3d_disc_kin_model}-\ref{sec.Cold_warm_molecular_phases}). Due to its intermediate ionization potential and high S/N, the \NeIIImu emission line is an optimal tracer of the ionised gas component within the dust-enshrouded inner disc. On the other hand, due to its higher ionization potential and the circumnuclear scales covered by the MIRI MRS observations, the \NeVmu is expected to trace gas that is highly ionised by the AGN radiation field. Finally, exploiting the multi-Gaussian fit of the \OIII and \NeIIImu emission lines described in Sects. \ref{sec.em_line_analysis_MUSE}-\ref{sec.em_line_analysis_MIRI} and shown in Fig. \ref{fig:fit_NeIII_H2S1_multigaussian_fit} we were able to disentangle the ionised gas component within the disc and the broader blue-shifted component, possibly associated with an AGN-driven wind.

\subsection{Multi-phase 3D disc kinematic model}\label{sec.3d_disc_kin_model}
\begin{figure*}
	\includegraphics[width=0.95\linewidth]{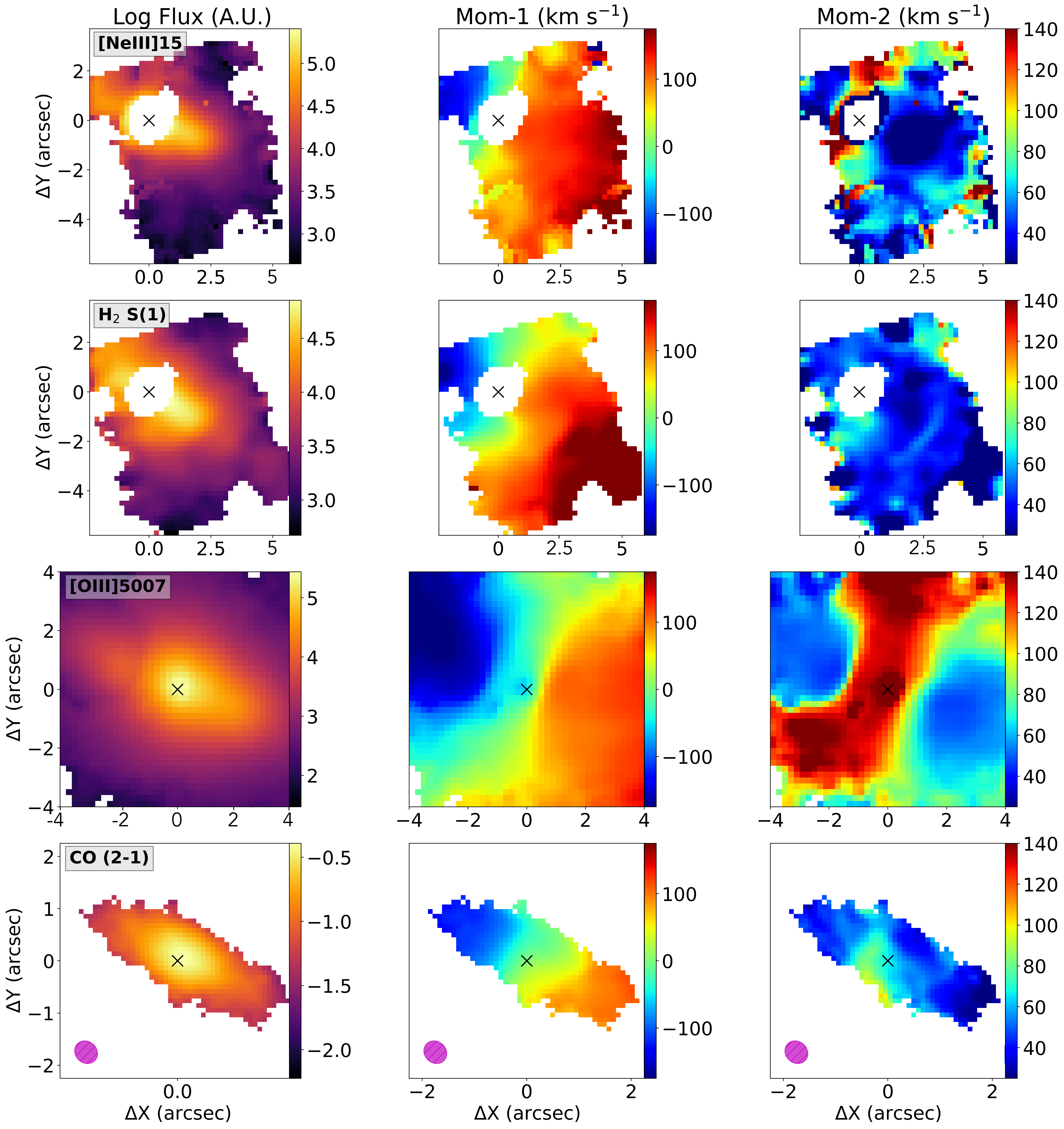}
    \caption{From top to bottom: NGC 424 disc emission traced by \NeIIImu, \hiisi, \OIII, and CO~(2-1) emission lines. The disc emission is traced by the total line profile of the \hiisi and CO~(2-1) transitions and by the narrower Gaussian component for the \NeIIImu and \OIII emission lines.  From left to right we show the emission line integrated flux, line of sight (LOS) velocity and velocity dispersion maps corrected for instrumental broadening. The integrated logarithmic flux maps are in unit of MJy/sr, 10$^{-20}$ erg/s/cm$^{-2}$, and Jy/beam for MIRI, MUSE, and ALMA maps, respectively. The ALMA beam is represented as a magenta oval. From top to bottom, a S/N mask of 3, 3, 5, and 3 was applied to moment maps. MIRI moment maps are obtained from the emission line fitting of the point-source subtracted data-cube.}
    \label{fig:disc_emission_4_tracers}
\end{figure*}

As a result of the multi-Gaussian fit described in Sect. \ref{sec.data_analysis}, we obtain an emission-line model cube representative of the disc component in NGC 424 by isolating the narrower Gaussian component used to reproduce the line profile in each spaxel. Figure \ref{fig:disc_emission_4_tracers} shows the disc component traced by the ionised, warm molecular and cold molecular emission, obtained from the MUSE, MIRI, and ALMA data-cubes, respectively. 
Overall, the moment maps in Fig. \ref{fig:disc_emission_4_tracers} show that the morphology of the rotating disc probed by different gas phases, is consistent across all tracers. We performed a detailed 3D kinematic modeling of the disc features to explore possible differences in the disc kinematics, infer the galaxy dynamical mass and explore the co-existence of multiple gas phases in the gaseous disc.

\begin{figure*}
	\includegraphics[width=0.9\linewidth]{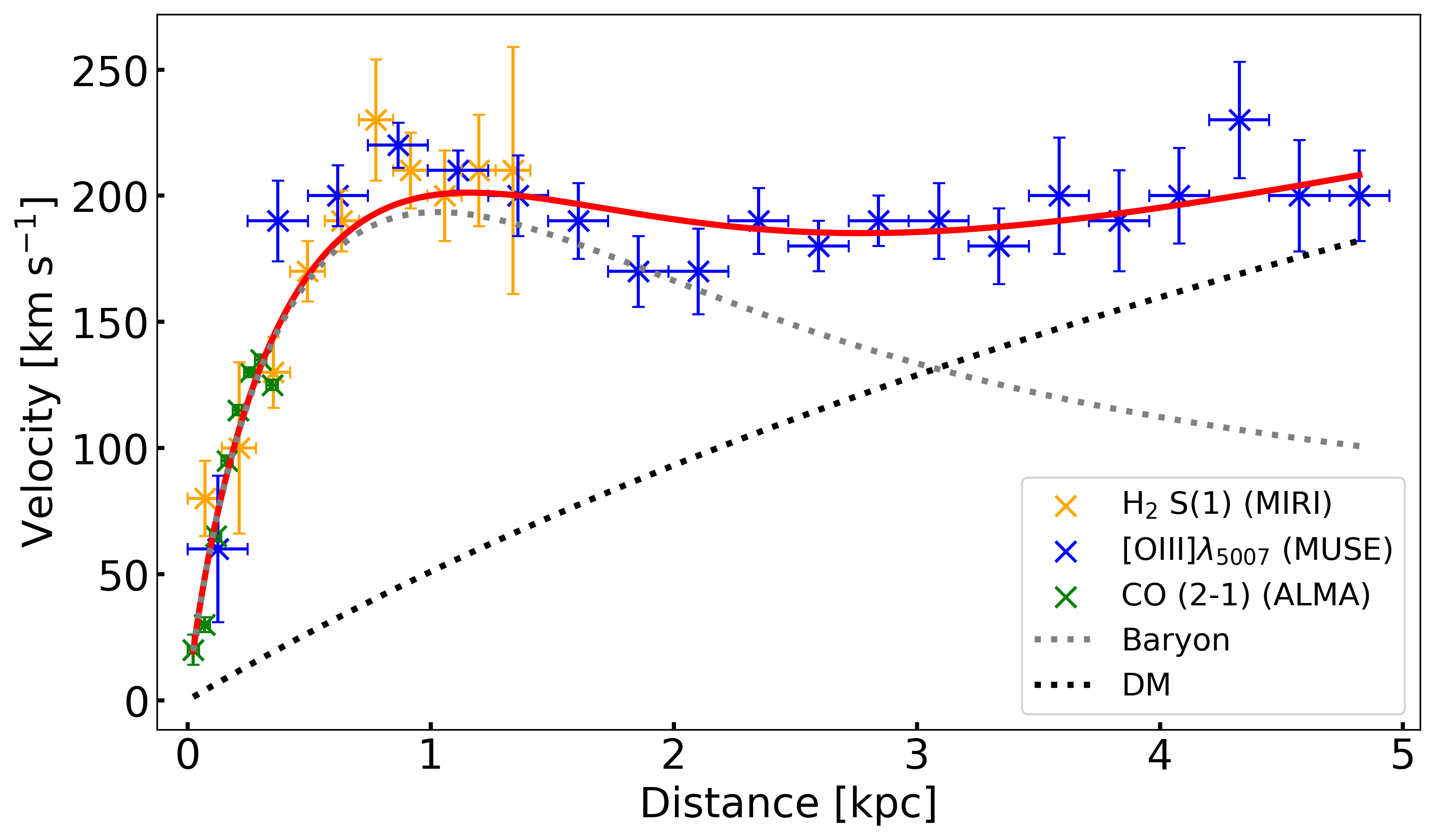}
    \caption{NGC 424 rotation curve inferred with \MOKA\ using the \hiisi (orange) and CO~(2-1) (green) emission, and the narrow component of the \OIII (blue) emission line, from MIRI/MRS, ALMA, and MUSE, respectively. The rotation curve obtained fitting the narrow component of the \NeIIImu shown in Fig. \ref{fig:disc_emission_4_tracers} is not reported, as the trend is similar to the \hiisi. The dotted black and grey curves represent the best-fit velocity profile of the dark matter (Eq. \ref{eq.dm_velocity}) and baryonic matter (Eq. \ref{eq.v_circ_final}), respectively. The total best-fit profile is shown as a solid red curve. The averaged rotating velocity and inclination values are listed in Tab. \ref{tab:moka_disc_fit}. Details on the \MOKA \ disc model are discussed in Sect. \ref{sec.3d_disc_kin_model}.}
    \label{fig:disc_v_profile}
\end{figure*}

To model the multi-phase disc properties we used our tool \MOKA (Modeling Outflows and Kinematics of AGN in 3D), presented in \citet{Marconcini2023}, and tailored to fit the multi-phase gas kinematic and geometry exploiting 3D data-cubes. We modeled the disc properties of the warm and cold molecular, and ionised gas phases using the disc component traced by the \hiisi, CO~(2-1), \NeIIImu,  and \OIII, respectively\footnote{As disc component we refer to the total line profile of the \hiisi and CO~(2-1) emission lines and to the narrower Gaussian component for the \NeIIImu and \OIII emission lines.}. For each gas phase we adopted a thin-disc geometry with a height smaller than the spatial PSF. We divided the disc in a different number of concentric circular shells, assuming as width of each shell the FWHM of the PSF. Therefore, the number of shells used to fit the observed disc properties in different gas phases in the final model depends on the IFS instrument tracing a specific gas phase. 

In each shell, we independently fitted a single free parameter, i.e. the disc circular velocity (V$_{rot}$), while keeping the intrinsic velocity dispersion of the model clouds and the disc inclination with respect to the line of sight ($\beta$) fixed. These assumptions are commonly adopted in kinematic modeling to mitigate degeneracies between the rotating velocity, velocity dispersion, and inclination \citep[e.g.][]{Forster_Schreiber2006,Epinat2010, Forster_Schreiber2018}, especially since we observe no kinematically disturbed motions as warps from the gas kinematic maps. To reduce the number of free parameters and thus reduce the degeneracies affecting the fit, the center of the disc and the position angle are not fitted with \MOKA \ but are measured with a separate routine. In particular, to infer the center of the disc model we collapsed the narrow component used to fit the line profile of each tracer and performed a 2D Gaussian fit to the collapsed image. Then we fixed the position angle tracing the disc minor axis from the measured centre. The fitting procedure can be summarized as follows. We initially fitted a single-shell model to the entire disc to derive global best-fit estimates of the circular velocity, disc inclination, and intrinsic velocity dispersion. Assuming that the derived inclination and velocity dispersion are constant across the disc, we then fixed these parameters and performed a multi-shell fit leaving the circular velocity free to vary with radius, and obtaining the rotation curve shown in Fig. \ref{fig:disc_v_profile}.  For completeness, we also checked that our measured value for the disc position angle from MUSE data is consistent with the values estimated with the \textit{PaFit} python package \citep{Krajnovic2006}, tailored to measure the global kinematic position angle from the gas kinematic in integral field data. The free and fixed best-fit parameters obtained for the disc fit of all the gas components using \MOKA \ are listed in Tab. \ref{tab:moka_disc_fit}. The comparison between the observed and modeled moment maps with \MOKA \ is described in App. \ref{sec_app_moka_fit}.

\input{mokafit_disc}

Figure \ref{fig:disc_v_profile} shows the disc rotation curve inferred with \MOKA \ fitting the disc component in MIRI, ALMA, and MUSE data. Interestingly, we observe that the disc velocity traced by different instruments is consistent at any distance up to the maximum FOV covered by each instrument. The gas circular velocity in the circumnuclear region traced by MIRI and ALMA has a low intrinsic and projected velocity, which slowly increases moving to larger distances, as traced by the fit of the MUSE data. The rotation curve shown in Fig. \ref{fig:disc_v_profile} is consistent with the expected profile for a simple gas disc in a galaxy, with no warps or distortions induced by a galactic bar. 

\subsection{Rotation curve fitting and dynamical mass}\label{sec.dynamical_mass_estimate}
From the inferred best-fit intrinsic rotating velocity profile in Fig. \ref{fig:disc_v_profile} we could estimate the dynamical mass of NGC 424 within the maximum radius covered by MUSE data. In particular, we assumed that the gas is distributed in a thin disc, that the stellar mass is distributed as the gas component with the gas mass surface density $\Sigma$(r) distributed as the surface brightness, and described as I(r) = I$_0$ $\times$ e$^{(-r / R_D)}$, with I$_0$ representing a normalisation constant.
Under these assumptions, \citet{Freeman1970} showed that the circular velocity can be expressed as:
\begin{equation}
    V(r)^2 = 4\pi G \Sigma_0 R_D r^2 \ ( I_0(r) K_0(r) -  I_1(r) K_1(r)),
\label{eq.v_circ_bessel}
\end{equation}
where $I$ and $K$ are the modified Bessel functions computed at $r =2R_D$. $\Sigma_0$ is a mass distribution constant used to normalize the gas and stellar contribution, and $R_D$ is the galaxy scale radius. The integrated mass surface density across all radii (M$_{dyn}$) can be expressed as M$_{dyn}$ = 2$\pi R_D^2 \Sigma_0$. Inserting the definition of M$_{dyn}$ in Eq.~\ref{eq.v_circ_bessel} we obtain:
\begin{equation}
    v(r)^2 = 2 (M_{dyn}/R_D) \ G r^2 \ ( I_0(r) K_0(r) -  I_1(r) K_1(r)).
\label{eq.v_circ_final}
\end{equation}

Additionally, since the observed rotation curve flattens after $\sim$ 1.5 kpc, we included the dark matter (DM) contribution to the rotation curve by assuming a Navarro-Frenk-White (NFW) profile \citep{Navarro1996, Navarro1997}, described as:

\begin{equation}
    \rm v_{dm}(r) = v_h \sqrt{\frac{r_0}{r} ln\left(\frac{r + r_0}{r}\right) - \frac{r_0}{r-r_0}},
\label{eq.dm_velocity}
\end{equation}

where $v_h$ = (4 $\pi$ r$_0^{2}$ $\rho_0$ G)$^{1/2}$ is the dark matter characteristic velocity, and $\rho_0$ and r$_0$ are the characteristic density and scale length of the dark matter profile, respectively. For the sake of simplicity, we followed the \cite{Lin2019} prescription and fitted $v_h$ instead of $\rho_0$.
Therefore, including the DM contribution to the observed rotation curve, we can write the total rotation velocity profile as: 
\begin{equation}
    \rm v(r) = \sqrt{v(r)^{2} + v_{dm}(r)^{2}},
\label{eq.tot_vrot}
\end{equation}
where $v(r)$ and $v_{dm}$ are defined in Eqs. \ref{eq.v_circ_final}-\ref{eq.dm_velocity}. 

Since we measured the intrinsic, de-projected multi-phase circular velocity profile using MIRI, MUSE, and ALMA data, covering scales from 20 pc with ALMA, up to 5 kpc with MUSE, we can fit Eq. \ref{eq.tot_vrot} to the rotation curve shown in Fig.\ref{fig:disc_v_profile}. Unfortunately, since our observations cover a maximum de-projected scale of $\sim$ 5 kpc from the nucleus we cannot properly constrain simultaneously the dark matter characteristic density and length, which are expected to contribute significantly at larger scales. Therefore, we assumed a characteristic scale length of r$_0$ = 8.1 $\pm$ 0.7 kpc as recently found by \citet{Lin2019} for the Milky Way and fit the dynamical mass ($M_{dyn}$), the baryonic mass scale radius ($R_D$) and the dark matter characteristic velocity $v_h$. As a result of the fit, we obtained the dashed red curve shown in Fig.~\ref{fig:disc_v_profile} which corresponds to the best-fit parameters $M_{dyn}$ = 1.09 $\pm$ 0.08 $\times$ 10$^{10}$ \msun, $R_D$ = 0.48 $\pm$ 0.02 kpc, and $v_h$ = 593 $\pm$ 35 \kms. Uncertainties on the derived parameters are evaluated at the 16th and 84th percentiles. The best-fit rotation curve, including the contribution from baryonic mass and dark matter, accurately reproduces the intrinsic circular velocity profile inferred with \MOKA. In particular, we observe that the dark matter contribution starts to be significant at radii larger than $\sim$ 2 kpc and that the NFW profile is well-suited to reproduce the observed rotation curve.

\subsection{3D ionised outflow kinematic modeling}\label{sec.3d_outflow_kin_model}
We used the \OIII and \NeIIImu emission lines, from MUSE and MIRI/MRS IFS data, to explore the ionised gas properties in the optical and MIR, respectively. As discussed in Sect. \ref{sec.em_line_analysis_MIRI} and \ref{sec.em_line_analysis_MUSE}, we used up to two Gaussian components to fit each observed emission line profile. Using a tailored number of Gaussian components to fit the observed emission line profiles is a well-tested method that has proven to be efficient in separating the emission originating from the rotating disc --typically observed as a narrow, symmetric component -- and the emission from outflowing gas -- observed as an asymmetric, broader component-- \citep{Venturi2018_magnum, Mingozzi2019}. In this section we discuss the ionised outflow properties as a result of the separation of symmetric (narrow) and asymmetric (broad) components used to reproduce the observed line profiles of the \OIII and \NeIIImu emission lines on a spaxel-by-spaxel basis.

To infer the ionised outflow properties observed both in \OIII and \NeIIImu emission lines we used again our \MOKA \ kinematic model which has proven to be particularly efficient in modeling outflow features at an unprecedented level of detail using IFS data from various instruments \citep[][Marconcini et al. 2025 accepted]{Marconcini2023, Cresci2023, Ulivi2024_arp, Perna2024_bal}.

In particular, we separately modeled the outflow properties traced by the broad components of the \OIII and \NeIIImu line profiles and finally discuss how the ionised outflow phase morphology and kinematic varies as a function of wavelength. 

To model the outflow properties we adopted a bi-conical-shaped outflow with a constant radial velocity field originating from the unresolved nucleus. To test the scenario of a tight interplay between the outflow and the elongated enhanced velocity dispersion region that will be described in Sect. \ref{sec_enhanced_vdisp_minor_axis}, we assumed the cone axis to be directed along the galaxy major axis, i.e. we fixed the outflow position angle to 118$^{\circ}$\footnote{The position angle is measured clockwise from North.}. 

\begin{figure*}
\centering
\includegraphics[width=0.7\linewidth]{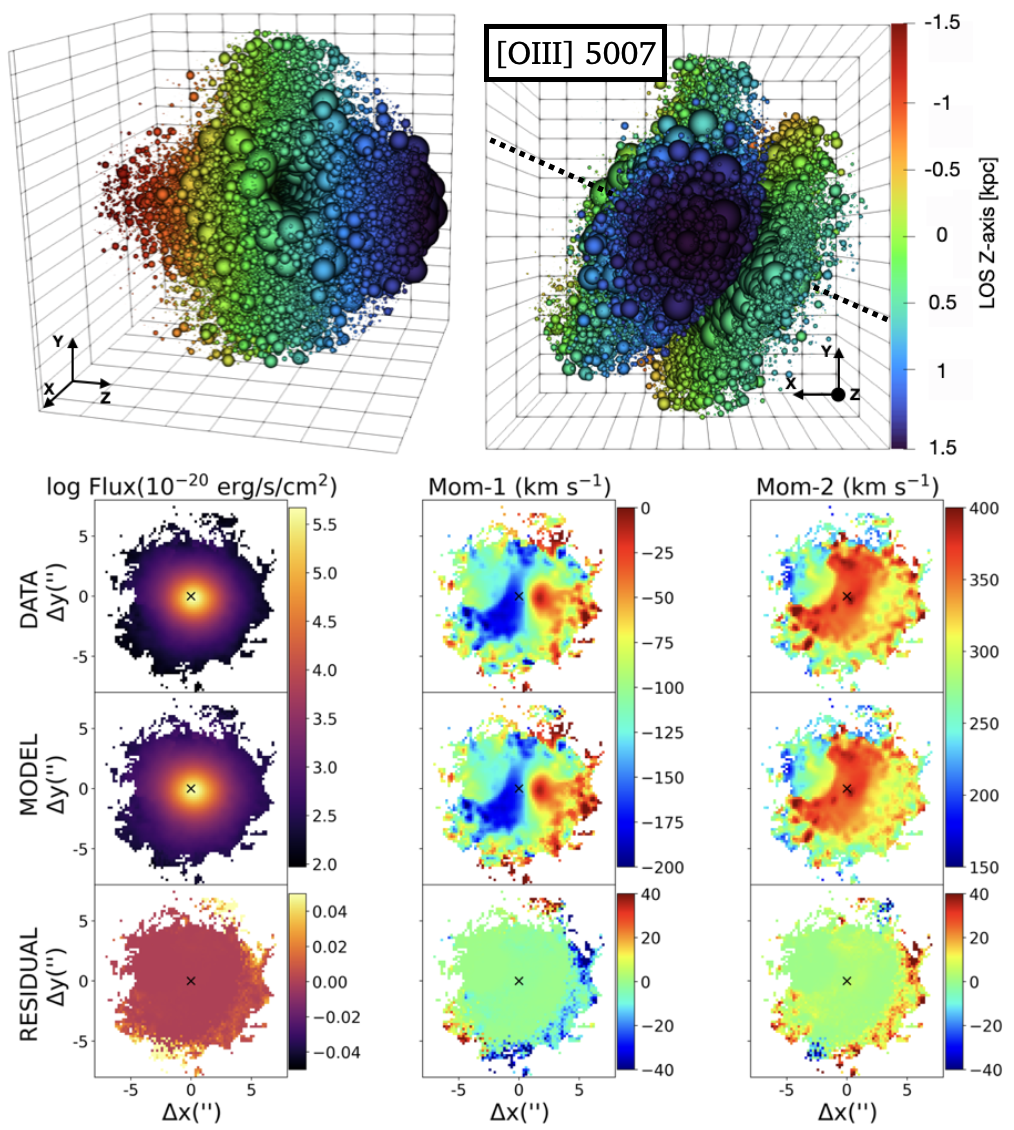}
    \caption{\MOKA \ best-fit model for the ionised outflow traced by the broad component of the \OIII from MUSE data. Top panels: 3D reconstruction of the ionised gas clouds color-coded based on their position w.r.t. the plane of the sky and observed along the outflow axis (left) and along the line of sight (right). The XY represents the plane of the sky, while the Z is the LOS. The dotted black lines in the right panel show the bi-conical outflow axis direction. According to the colorbar, blue and red clouds are blue-shifted and red-shifted, respectively. The bubble size is representative of the intrinsic cloud flux. Bottom panels: Data (top), \MOKA \ best-fit (middle) and residual (bottom) moment maps, i.e. the integrated flux, the LOS velocity, and velocity dispersion maps. The residual maps are obtained by subtracting the model from the observed moment maps. Maps are masked at a S/N of 3.}
    \label{fig:outflow_moka_OIII}
\end{figure*}

For the MUSE data, since the observed outflowing gas clouds extend up to 5 \arcsec in projection (i.e., 1.2 kpc) and we detected extended blue-shifted emission towards the East direction, (see bottom panels in Fig. \ref{fig:outflow_moka_OIII}) we assumed a large outer opening angle of 70$^{\circ}$. A wide outer opening angle is necessary to simultaneously reproduce such extended blue-shifted outflow features on such a compact scale. Similarly, towards the West direction, we observe a more collimated red-shifted wind which is probably obscured by the galactic disc and represents the counterpart of the blue-shifted wind on the East side. Therefore, to model the observed features traced by the \OIII emission line we adopted a bi-conical geometry. 

The free parameters of the model are the outflow inclination, the intrinsic radial velocity, velocity dispersion, and the inner opening angle of the cone. As a result of the \MOKA \ fit, we obtained an inclination of the cone axis with respect to the plane of the sky of $\beta$ = 35$^{\circ} \pm$ 3$^{\circ}$, an outflow radial velocity of $V_{out}$ = 990 $\pm$ 35 km s$^{-1}$, an intrinsic outflow velocity dispersion of $V_{disp}$ = 95 $\pm$ 25 km s$^{-1}$, and inner opening angle of 20$^{\circ} \pm$ 7$^{\circ}$ (see also Tab. \ref{tab:moka_outflow_fit}). 

Based on the best-fit outflow parameters and taking into account the disc inclination (see Sect. \ref{sec.3d_disc_kin_model} and Tab. \ref{tab:moka_disc_fit}), we remark that the outflow is directly impacting on the gaseous disc, having the outflow and disc a similar inclination with respect to the line of sight. Moreover, based on the \MOKA \ best-fit the outflow has an inner cavity surrounding its axis, which can be ascribed to the fact that it cannot propagate freely while interacting with the disc, due to the high-resistance path and column density. 

\begin{figure*}
\centering
\includegraphics[width=0.7\linewidth]{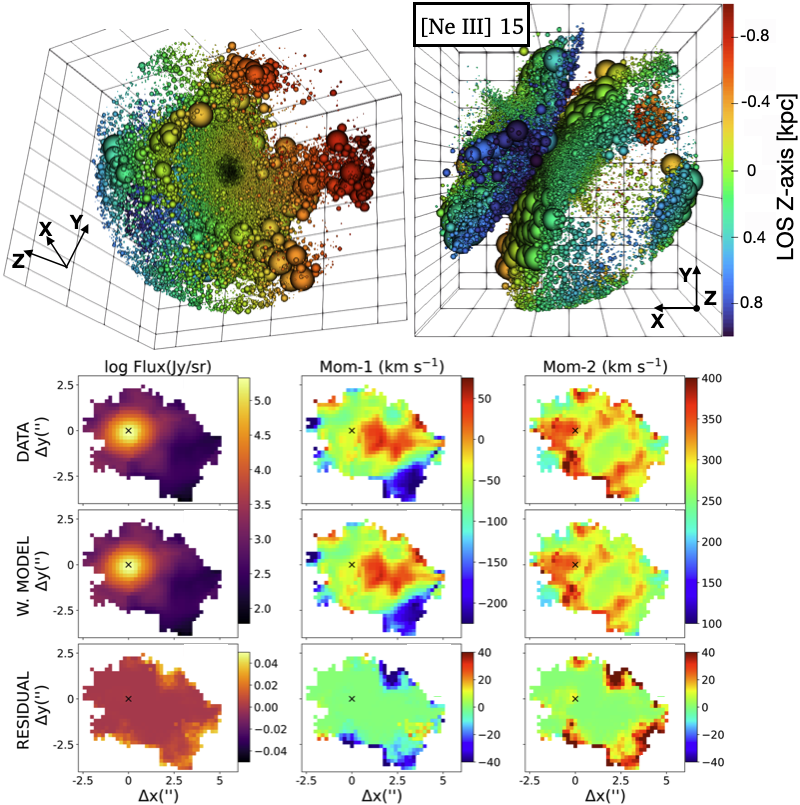}
    \caption{Same as in Fig. \ref{fig:outflow_moka_OIII} for the \NeIIImu emission line from MIRI MRS Ch3. Maps are masked at a S/N of 2.}
    \label{fig:outflow_moka_NeIII}
\end{figure*}

To infer the outflow properties traced by the Mid-IR \NeIIImu emission line, we followed a similar approach as for the \OIII emission line. We considered the broad Gaussian component of the \NeIIImu line profile in each spaxel and created the moment maps shown in Fig. \ref{fig:outflow_moka_NeIII}, which due to the low S/N of the broad component of the \NeIIImu trace only the West direction. Interestingly, since \NeIIImu traces emission at longer wavelengths with respect to the \OIII emission line and is thus less obscured by the galactic disc, we detect the red-shifted side of the NW ionisation cone at much higher S/N.

To model the \NeIIImu\ outflow we assumed a conical-shaped wind morphology with the same inner and outer opening angle and position angle as for the \OIII-traced outflow. Using \MOKA \ to reconstruct the outflow morphology and kinematics we estimated an outflow inclination for the West side of $\beta$ = 25$^{\circ} \pm$ 7$^{\circ}$ with respect to the plane of the sky, an outflow radial velocity of $V_{out}$ = 975 $\pm$ 30 km s$^{-1}$, an intrinsic outflow velocity dispersion of $V_{disp}$ = 90 $\pm$ 10 km s$^{-1}$, consistently with the \OIII estimates (see Tab. \ref{tab:moka_outflow_fit}). Interestingly, as reported in Tab. \ref{tab:moka_outflow_fit}, we found consistent estimates for the \OIII and \NeIIImu outflow intrinsic kinematic and inclination, highlighting the pivotal role of a sophisticated kinematic model as \MOKA \ to account for the complex outflow features.

\input{mokafit_outflow}

\subsection{Highly ionised outflow}\label{sec.highly_ionised_outflow}
\begin{figure}
	\includegraphics[width=\linewidth]{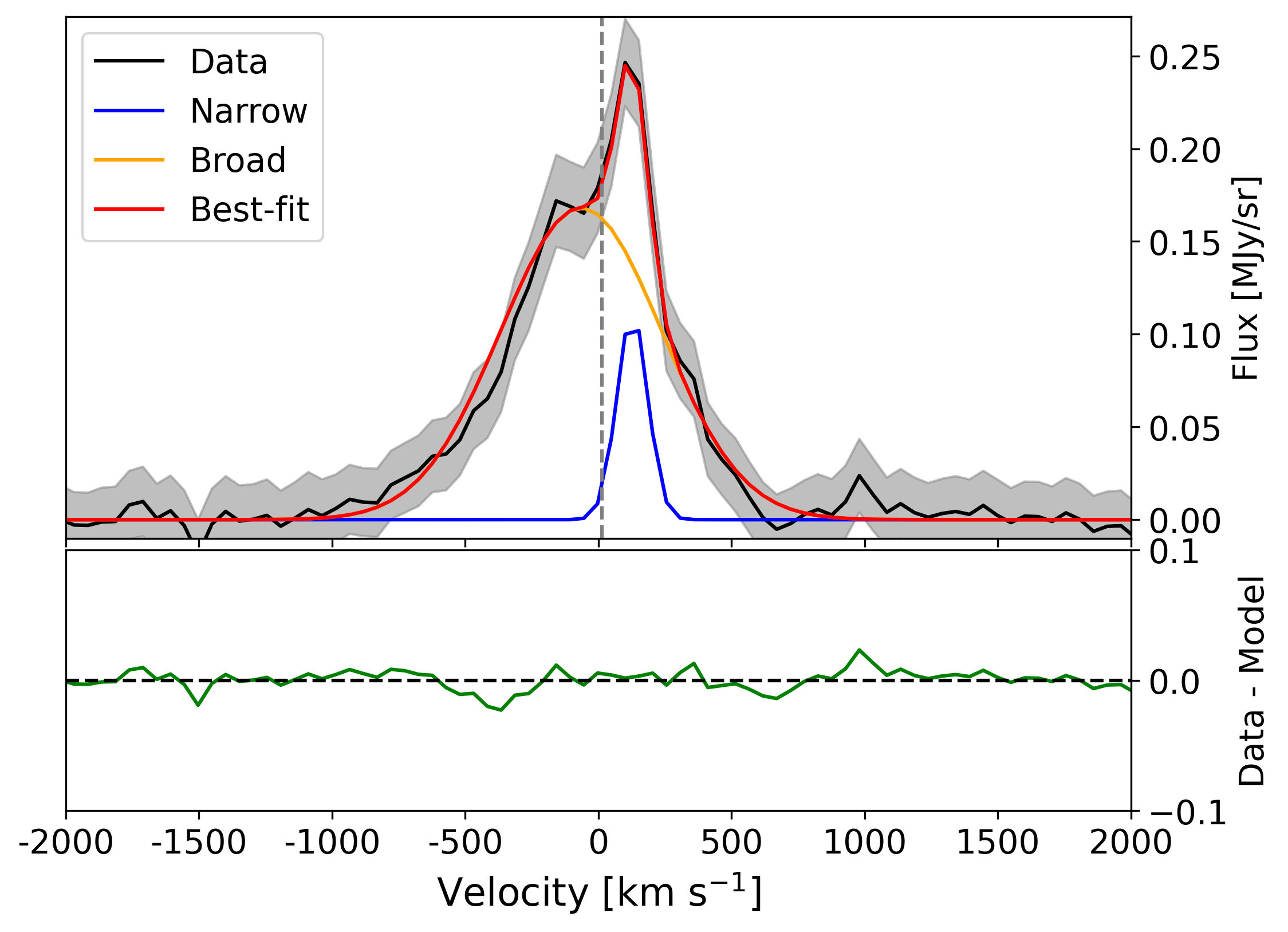}
    \caption{Multi-Gaussian fit of the integrated \NeVmu emission line from the total MIRI/MRS Ch3 MEDIUM band. Data and total best-fit model are in black and red, respectively. The narrow and broad Gaussian components are shown as solid blue and orange lines, respectively. The bottom panel shows the residuals in green. The vertical gray dashed line marks the rest-frame wavelength of the \NeVmu transition.}
    \label{fig:NeVfit}
\end{figure}

The highest S/N highly ionised emission line in our MIRI/MRS data is the \NeVmu that falls in the Ch3 MEDIUM band. Unfortunately, despite the spatial and spectral subtraction of the PSF, the S/N on the broad component of the \NeVmu emission line is not sufficient to provide a reliable constraint on the highly ionised wind morphology. Therefore, we integrated the emission-line profile of the \NeVmu over the entire FOV and performed a two Gaussian-component fit to the line profile, which presents the same blue-shifted line wing as for the \NeIIImu and \OIII line profile (see Fig. \ref{fig:NeVfit}). As a result of the fit we found a velocity dispersion for the broad component of 283 $\pm$ 17 \kms and a blue-shift with respect to the rest-frame emission of 61 $\pm$ 5 \kms. Assuming that the \NeVmu-traced outflow has the same morphology as the \OIII- and \NeIIImu-traced outflows and that the outflow is propagating at constant radial velocity, we can correct the maximum observed \NeVmu-traced outflow velocity of 812 $\pm$ 47 \kms for projection effects\footnote{This value is estimated from the maximum value between the first and 99-th percentile of the line profile. This method to estimate the intrinsic outflow velocity, not corrected for projection effects, has been deeply discussed in \citet{Marconcini2023}.}. Indeed, assuming the same inclination of the \NeIIImu outflow of 25$^{\circ}$ $\pm$ 7$^{\circ}$ (see Tab. \ref{tab:moka_outflow_fit}) we found V$_{out, \NeVmu}$ = 900 $\pm$ 50 \kms. Such value for the intrinsic outflow velocity inferred for the highly-ionised phase is consistent with the values derived for the warm ionised phase listed in Tab. \ref{tab:moka_outflow_fit}.

\subsection{Ionised outflow energetics}\label{sec.outflow_energetic}
To infer the energetic impact of the ionised outflow in NGC 424 we exploited the tomographic outflow reconstruction obtained with \MOKA, which provides the distribution of ionised gas clouds within the outflow and the intrinsic (i.e., de-projected) outflow properties. In particular, similarly to the method used in \citet{Marconcini2023} and based on the model by \citet{CanoDiaz2012}, we computed the amount of ionised mass traced by the \OIII emission lines in each spaxel using the relation:
\begin{equation}
    \rm M_{out} = 5.33 \times 10^7  \left( \frac{L_\text{[OIII]}}{10^{44} \, \text{erg} \, \text{s}^{-1}} \right) \left( \frac{n_e}{1000 \, \text{cm}^{-3}} \right)^{-1} \ \ M_\odot
    \label{eq.m_out}
\end{equation}
\noindent

where $L_{\rm [OIII]}$ is the luminosity of the broad component of the \OIII emission line corrected for dust attenuation (see App. \ref{app_sec_electron_density}), $\rm n_e$ is the average electron density of the ionised gas. Therefore, using Eq. \ref{eq.m_out} we can convert the flux density of the ionised wind in mass. Assuming the flux density within each spaxel to be constant over time and the outflow to subtend a solid angle $\Omega$ we can use Eq. \ref{eq.m_out} to compute the mass outflow rate ($\rm \dot M_{out}$), i.e. the amount of ionised mass crossing a distance R with velocity V$_{out}$, as:
\begin{equation}
    \rm \dot M_{out} = \frac{M_{out} V_{out}}{R} 
    \label{eq.m_out_rate}
\end{equation}
\noindent

In order to be able to compare the outflow energetics traced by the \OIII and Mid-IR emission lines, and due to the limited FOV of MIRI with respect to MUSE, in the following we will refer to the \OIII luminosity extracted from the same region considered for the \NeIIImu and \NeVmu emission lines, i.e. the MIRI Ch3 FOV.

For the ionised phase traced by the \OIII emission line we found a total outflowing gas mass of $\rm M_{out}$ = 24 $\pm$ 11 $\times$ 10$^4$ \msun using an average electron density of $\rm <n_{e}>$ = 570 $\pm$ 360 cm$^{-3}$ obtained from the total line profile of the \SIIall doublet of MUSE data (see App. \ref{app_sec_electron_density} and Fig. \ref{fig:eletron_density_muse}). Using the best-fit values for the outflow velocity of V$_{out}$ = 990 $\pm$ 35 km s$^{-1}$ and the outflow de-projected size inferred from MIRI observations of 1.1 $\pm$ 0.1 kpc (assuming as uncertainty half the FWHM of the MUSE PSF) we estimated a mass outflow rate of $\rm \dot M_{out}$ = 2.2 $\pm$ 1.0 $\times$ 10$^{-1}$ \msun yr$^{-1}$. As a comparison, we computed the outflowing gas mass and mass outflow rate from the broad component of the \Halpha recombination line assuming the same gas electron density as for the \OIII analysis. After  correcting for extinction we found $\rm M_{out}$ = 90 $\pm$ 46 $\times$ 10$^4$ \msun and $\rm \dot M_{out}$ = 8.8 $\pm$ 4.5 $\times$ 10$^{-1}$ \msun yr$^{-1}$. Therefore, comparing the outflowing gas masses traced by \OIII and \Halpha we found $\rm M_{out, \OIII} / M_{out, \Halpha} = 0.26$. This value is consistent with the typical ratio measured in Seyfert galaxies and such discrepancy has to be ascribed to the different volume of 3D regions of \OIII and \Halpha emission \citep[see also][]{Carniani2015, Karouzos2016}. Indeed, the volume of gas emitting the \Halpha recombination line is expected to be larger and more extended along the line of sight with respect to the \OIII line, thus tracing larger amount of ionised mass. 

To infer the energetics of the ionised outflow traced by the Mid-IR emission lines, such as the \NeIIImu and \NeVmu, we need to take into account the Neon emissivity as well as its abundance. A detailed description of the procedure to infer the amount of ionised mass traced by Mid-IR emission line will be presented in Ceci et al. 2025 (in prep.). Assuming a solar abundance of [Ne/H] of 7.93 \citep{Asplund2009}, a ionised gas temperature of T$_e$ = 10$^4$ K we can write the ionised mass traced by the Neon emission lines as:

\begin{equation} \label{eq.outflow_mass_neon}
    \rm M_\text{Neon} = \alpha \,  
    \left( \frac{L_\text{Neon}}{10^{36} \, \text{erg} \, \text{s}^{-1}} \right) 
    \left( \frac{n_e}{200 \, \text{cm}^{-3}} \right)^{-1}  \ M_\odot
\end{equation}

\input{energetics}

where $\alpha$ is a factor that takes into account the emissivity of different transition of the Neon atom\footnote{Using $Pyneb$ \citep{Luridiana2012, Luridiana2015} and a gas electron density of 570 $\pm$ 360 cm$^{-3}$ we computed a lines emissivity of $\gamma_\text{\NeIIImu} = 1.25 \times 10^{-21}$ erg s cm$^{-3}$ and $\gamma_\text{\NeVmu} = 1.93 \times 10^{-20}$ erg s cm$^{-3}$ for the \NeIIImu and \NeVmu emission lines, respectively.}. In particular, we get $\alpha$ = 46 and $\alpha$ = 3, for the \NeIIImu and \NeVmu emission lines, respectively.

Therefore, we estimated outflowing gas mass of $M_\text{\NeIIImu}$ = 5.4 $\pm$ 2.7 $\times$ 10$^4$ \msun and $M_\text{\NeVmu}$ = 1.9 $\pm$ 0.9 $\times$ 10$^4$ \msun. Using Eq. \ref{eq.m_out_rate} and the outflow kinematic properties listed in Tab. \ref{tab:moka_outflow_fit} we estimated a mass outflow rate of $\rm \dot M_{out,\NeIIImu}$ = 4.5 $\pm$ 2.3 $\times$ 10$^{-2}$ \msun yr$^{-1}$ and $\rm \dot M_{out,\NeVmu}$ = 1.6 $\pm$ 0.8$\times$ 10$^{-2}$ \msun yr$^{-1}$. 

Finally, we compute the outflow kinetic energy (E$_{kin}$), kinetic luminosity (L$_{kin}$) and momentum rate ($\dot p_{out}$) using the following expressions: $\rm E_{out} = \frac{1}{2} M_{out} V_{out}^2, L_{kin} = \frac{1}{2} \dot{M}_{out} V_{out}^2, \dot p_{out} = \dot{M}_{out} v_{out}$.

The results for the outflow energetics traced by the \NeVmu, \NeIIImu, and \OIII emission lines are listed in Tab. \ref{tab:energetics}. To estimate the ionised outflow energetic traced by the \OIII emission line, we corrected the line flux for the dust attenuation, using the resolved dust attenuation map shown in Fig. \ref{fig:eletron_density_muse} computed assuming a \citet{Calzetti2000} attenuation law (see also App. \ref{app_sec_electron_density}). To correct the Mid-IR emission line fluxes, we used the \citet{Gordon2023} attenuation curve, tailored to the MIR regime.
The estimates for the ionised outflowing gas mass as well as the mass outflow rate traced by the \OIII emission line are largely dominant with respect to the \NeIIImu and \NeVmu estimates. In Sec. \ref{sec_discussion_multiphase_winds} we present a possible interpretation for this result as well as a comparison with estimates from the literature.


Overall, our analysis exploiting archival \OIII observations and recent MIRI/MRS observations, revealed that in NCG 424 the largest contribution to the outflowing wind energetics is carried by the warm ionised gas traced by the \NeIIImu and \OIII emission lines. To a lesser extent, the highly-ionised gas traced by \NeVmu emission line reveals a mass outflow rate an order of magnitude smaller with respect to the warm-ionised phase. 

\subsection{Enhanced velocity dispersion along the galaxy minor axis}\label{sec_enhanced_vdisp_minor_axis}
The narrow component velocity dispersion maps of \NeIIImu, CO~(2-1), \OIII, and to a lesser extent of \hiisi, show an elongated enhancement of gas velocity dispersion along the disc minor axis (see Fig.~\ref{fig:disc_emission_4_tracers}). Moreover, perpendicularly to such enhancement we observe deep minima of the gas velocity dispersion, especially in the ionised phase. Such features resemble the enhancement of gas velocity dispersion perpendicular to the outflow and radio jet direction in local Seyfert galaxies observed in previous works \citep[e.g.][]{Gonzales_delgado2002, Schnorr_muller2014, Schnorr_muller2016, Freitas2018, Finlez2018, Shimizu2019, Durre_mould2019, Shin2019, Feruglio2020, Venturi2021_turmoil, Girdhar2022, Ulivi2024_sigmaperp}. The most credited scenario for this phenomenon suggests that such enhancement is due to the impact of the radio jet onto the galaxy disc, perturbing the ambient gas in the ionised phase. Indeed, the observed extended velocity dispersion perpendicular to AGN jets and ionization cones in sources with a low jet-disc inclination might result from a strong impact of the jet on the host material. Such scenario is also consistent with cosmological simulations, showing that radio jets are capable of affecting and altering the host galaxy equilibrium \citep{Weinberger2017, Pillepich2018} and with jet-ISM interaction hydrodynamic simulations \citep{Mukherjee2016, Mukherjee2018, Meenakshi2022}. 

Such phenomenon is observed also in the H$_2$ warm molecular \citep{Riffel2014, Riffel2015, Diniz2015} and CO~(2-1) cold molecular components \citep{Shimizu2019}. Studies presenting these results have proposed different explanations for such phenomenon, all of which invoke a tight interaction among the AGN, the radio jet, and the host galaxy, discarding the hypothesis of instrumental effects such as beam smearing as the cause.  

\begin{figure}
	\includegraphics[width=\linewidth]{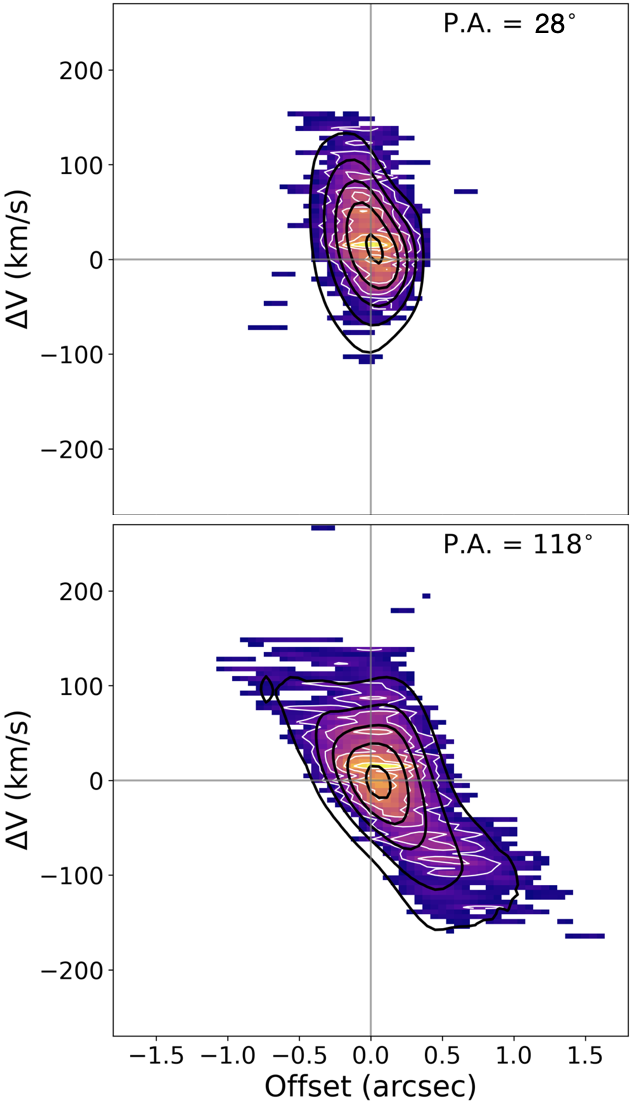}
    \caption{Position-Velocity diagrams extracted from a slit of width 0.8\arcsec along the NGC 424 CO~(2-1) kinematic minor axis (top panel, P.A.=28$^{\circ}$, see Tab. \ref{tab:moka_disc_fit}) and major axis (bottom panel, P.A.=118$^{\circ}$). CO~(2-1) emission is masked at S/N $\leq$ 3. White contours correspond to arbitrary levels of the observed emission. Black contours correspond to arbitrary levels extracted from the best-fit model cube obtained with \MOKA\ shown in Fig. \ref{fig:MOKA_CO21_momamps} and described in Sect. \ref{sec.3d_disc_kin_model}.}
    \label{fig:PVD}
\end{figure}

All the aforementioned works found the velocity dispersion to be enhanced perpendicularly to the host galaxy radio jets and ionization cones. Unfortunately, there is only weak evidence of a radio-jet in NGC 424. Indeed, \citet{Nagar1999} observed a slightly elongated E-W radio structure using a VLA 20 cm image, later on confirmed by \citet{Mundell2000} with higher-resolution VLA 3 cm observations. Due to the low sigma detection of such radio detections, the presence of a radio jet is still unconfirmed. Nevertheless, we investigated the presence of such radio jet from VLA 8.49 GHz (X band) data\footnote{The VLA data presented in Fig. \ref{fig:disc_emission_4_tracers} are the same data presented in \citet{Mundell2000}, re-analysed in 2009 and downloaded from the NRAO archive \url{https://www.vla.nrao.edu/astro/nvas/avla.shtml}} and found no evidence of bright radio spots or E-W structures, therefore casting doubts on the presence of a radio jet in NGC 424.. 

In Secs. \ref{sec.em_line_analysis_MUSE} and \ref{sec.em_line_analysis_MIRI} and in Fig. \ref{fig:fit_NeIII_H2S1_multigaussian_fit} we showed the presence of broad blue-shifted emission possibly associated with an AGN-driven wind which we further confirmed in Secs. \ref{sec.3d_outflow_kin_model} and \ref{sec.highly_ionised_outflow}. In particular, in Sect. \ref{sec.3d_outflow_kin_model} we showed that such wind is well-fitted by a large opening-angle (bi)conical outflow propagating at velocities of $\sim$ 10$^{3}$ \kms and interacting with the host galaxy due to the low inclination angle of the outflow axis with respect to the galaxy plane. Thanks to our detailed kinematical modeling of the AGN-driven wind, mainly traced by \NeVmu and \OIII emission, we strengthen the scenario of significant outflow-disc interaction possibly causing the observed enhancement of gas velocity dispersion along the minor axis, despite we have no evidences of a radio-jet in NGC 424, which instead is a crucial ingredient in the scenario proposed by \citet{Venturi2021_turmoil} and \citet{Ulivi2024_sigmaperp}.

An alternative scenario is discussed by \citet{Pilyugin2021a} as such elongated enhancement of the gas velocity dispersion along the galaxy minor axis has been observed also in a sample of MaNGA galaxies. The authors suggest that the presence of the region of enhanced gas velocity dispersion can be an indicator of a specific evolutionary path or evolution stage of the host galaxy. Moreover, they do not find evidence for such region to be related with the presence of an AGN and thus to a radio-jet, but can  rather be explained with a radial component of the gas velocity dispersion larger than the azimuthal and vertical components. Nevertheless, a complete understanding of the cause of this higher velocity dispersion in the radial direction is still missing.

\subsection{Cold and warm molecular phases}\label{sec.Cold_warm_molecular_phases}
We used the MIRI/MRS Ch3 LONG data-cube and the ALMA band 6 high-resolution data cube to explore the warm and cold molecular gas phase properties, as traced by the \hiisi and CO~(2-1) transitions (see Secs. \ref{sec.em_line_analysis_MIRI} and \ref{sec.em_line_analysis_ALMA}).

\begin{figure*}
	\includegraphics[width=\linewidth]{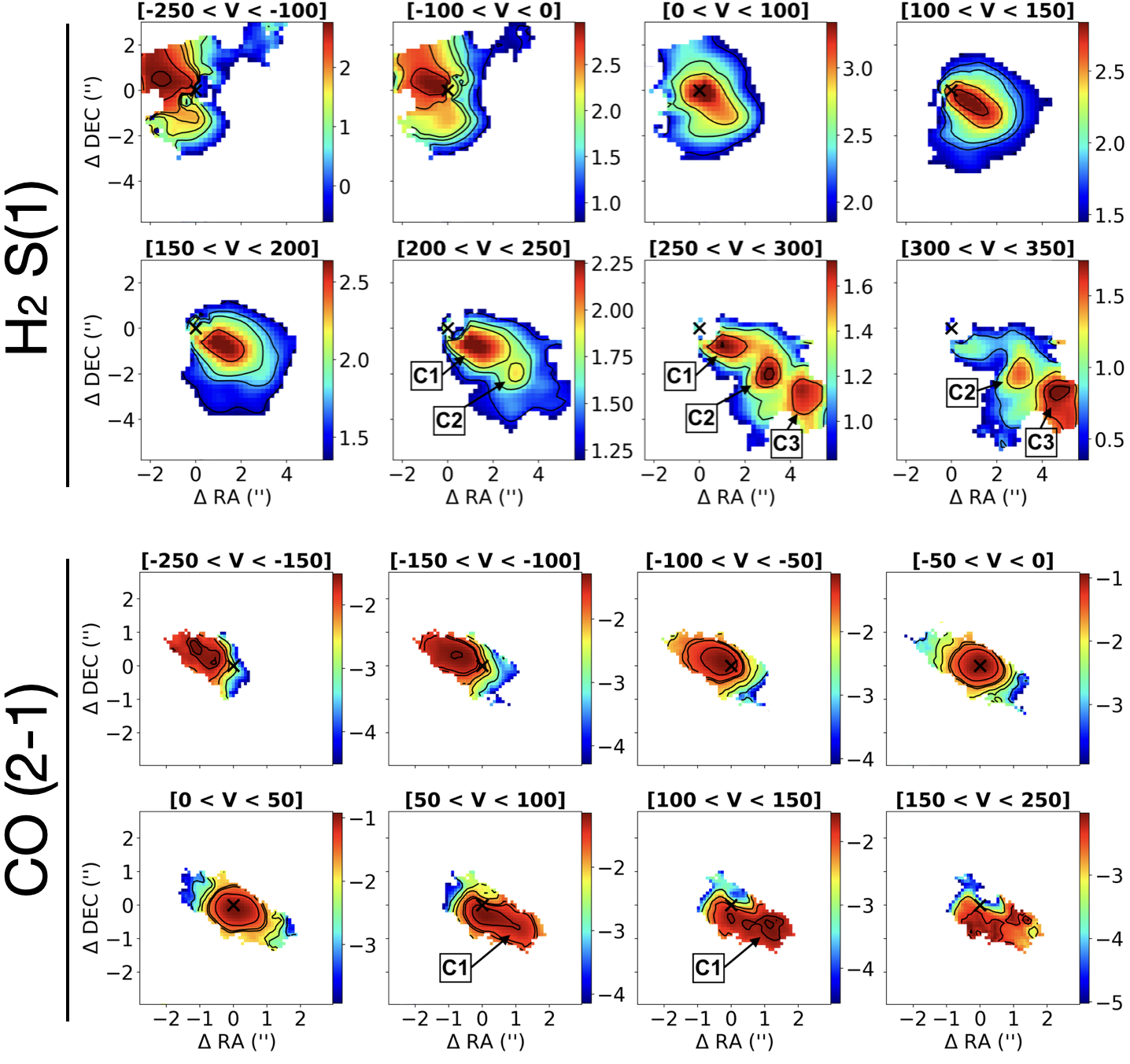}
    \caption{Velocity channel maps at different velocity bins for the warm (\hiisi, top panels) and cold molecular (CO~(2-1), bottom panels) gas components. Channel maps are in logarithmic scale and in units of MJy/sr (top panels) and Jy/beam (bottom panel). Contours represent arbitrary flux level. Molecular gas knots in the galactic disc, discussed in Sect. \ref{sec.Cold_warm_molecular_phases}, are labelled as C1, C2, and C3.}
    \label{fig:h2_co21_velchan_maps}
\end{figure*}

To investigate possible non-circular motions in the cold molecular phase, we computed Position-Velocity Diagrams (PVD) along the major and minor axis of the galaxy, from a 0.8\arcsec slit. As shown in Fig. \ref{fig:PVD}, the pure circular motions modelled with \MOKA (see Fig. \ref{fig:MOKA_CO21_momamps} and Sect. \ref{sec.3d_disc_kin_model}) and shown in black contours, reproduce the observed features, with no evidence of outflowing gas either along the major or minor axis, as also suggested by the residual maps shown in bottom panel in Fig.~\ref{fig:MOKA_CO21_momamps}. Exploiting PVD to infer the possible presence of outflowing material from ALMA data is a well-tested method that here confirms the pure rotating nature of the cold molecular phase in NGC 424 \citep{RamosAlmeida2022, Audibert2023, Zanchettin2025}.
Moreover, Fig. \ref{fig:h2_co21_velchan_maps} shows the velocity channel maps for the warm and cold molecular gas phases, traced by \hiisi and CO~(2-1) emission lines, respectively. As suggested by the moment maps (Fig. \ref{fig:disc_emission_4_tracers}) and line profiles (Fig. \ref{fig:fit_NeIII_H2S1_multigaussian_fit}), the velocity channel maps confirm the absence of outflowing gas in the cold and warm molecular phases, as suggested by the non detection of emission in the high-velocity channels. A more quantitative assessment of the non-detection can be made by converting the RMS noise of the spectrum computed over the high-velocity regions of the line (e.g. here we compute it in both the blue-shifted and red-shifted 250-500 \kms spectral region) into a molecular gas mass sensitivity. In particular, we extracted the spectrum from a circular region centred on the nucleus and with a radius equal to the ALMA FWHM (i.e., $\sim$ 0.45\arcsec). Then, assuming a CO~(2-1) to CO~(1-0) luminosity ratio of R$_{21}$ = 0.7 \citep{Leroy2009, Sandstrom2013}, and a conservative CO-to-H$_2$ conversion factor $\alpha_{CO}$ = 4.3 \msun pc$^{-2}$ (K km s$^{-1}$)$^{-1}$ \citep{Bolatto2013}, we estimate a 1 $\sigma$ molecular gas mass upper limit of 3.3 $\times$ 10$^{6}$ \msun. As a comparison, this upper limit on the outflow mass is $\leq$ 10\% of the total molecular gas mass estimated from the same region. This fraction is consistent with findings in local galaxies, where the molecular outflow component is often found to be only a small fraction of the total molecular reservoir \citep{Cicone2014, Fluetsch2019, Lamperti2022}.

Interestingly, channel maps highlight the combination of diffuse and extremely clumpy structures along the disc major axis direction. In particular, flux maps in velocity intervals between [-200, 150] km $^{-1}$ and [-150, 50] km $^{-1}$ for the \hiisi and CO~(2-1) emission lines respectively, appear smooth and diffuse. On the other hand at projected velocities larger than 150 km $^{-1}$ and 50 km $^{-1}$ for the \hiisi and CO~(2-1) emission lines respectively, we observe well-defined sub-structures, not observed in any of the ionised gas flux maps. Indeed, exploiting the MIRI/MRS spatial resolution after subtracting the PSF we noticed three resolved main clumps (C1, C2, C3) separated by a projected distance of 460 pc in the three most red-shifted channel maps of the \hiisi emission line shown in Fig. \ref{fig:h2_co21_velchan_maps}. Similarly, a clumpy structure is also detected at blue-shifted velocities in the \hiisi and CO~(2-1) channel maps. Interestingly, the C1 clump detected in \hiisi is also highlighted in the [50, 100] \kms CO~(2-1) channel map. 

The co-spatiality of the same resolved clump (i.e., C1) both in \hiisi and CO~(2-1), observed at different projected velocities suggest a stratification of the molecular gas where the emission of the colder phase (i.e., CO~(2-1)) originates from a more embedded region with respect to the warm phase (i.e., \hiisi). The same conclusion is also supported by the lower circular velocity of the cold molecular phase as inferred from the kinematic fit in Sect. \ref{sec.3d_disc_kin_model} (see also Tab. \ref{tab:moka_disc_fit}).


\section{Discussion}\label{sec.discussion}

\subsection{Multi-phase elongated enhancement of the velocity dispersion}
As shown in Fig. \ref{fig:disc_emission_4_tracers}, we found remarkable evidence of enhanced gas velocity dispersion along the galaxy minor axis both in the ionised and molecular phases, cold and warm, as traced by \NeIIImu, \OIII, \hiisi, and CO~(2-1) emission, respectively. Moreover, such enhancement appears to be perpendicular to the compact ionised outflow detected both in \NeIIImu, \NeVmu, and \OIII transitions. Thanks to a detailed modeling of the outflow properties (see Figs. \ref{fig:MOKA_OIII_momamps} and \ref{fig:MOKA_NeIII_momamps}, and Sect. \ref{sec.3d_outflow_kin_model}) we discovered that the outflow is impacting the galactic disc and is well-reproduced by a conical wind with an inner cavity along the axis. 


Up to date, no previous works detected such phenomenon of enhanced velocity dispersion in the ionised, warm and molecular gas phases simultaneously. 
Moreover, despite previous AGN-focused observations and simulations ascribe the observed enhancement of velocity dispersion (see Sect. \ref{sec_enhanced_vdisp_minor_axis}) to the radio-jet impacting onto the host galaxy disc \citep{Mukherjee2016, Venturi2018_magnum, Talbot2022, Audibert2023, Ulivi2024_sigmaperp}, we found the same pattern of velocity dispersion enhancement, with no evidence of a radio jet, consistently with the findings of \citet{Pilyugin2021a}. In the following section we present a detailed 3D morphological and kinematical analysis of the wind suggesting that the ionised outflow might play a crucial role in disturbing the ambient material and thus contributing to the observed enhancement of the gas velocity dispersion, despite the absence of a radio-jet. Finally, since we do not find evidence of outflowing molecular gas but nevertheless observe enhanced velocity dispersion along the galaxy minor axis (see \hiisi and CO~(2-1) panels in Fig. \ref{fig:disc_emission_4_tracers}) we suggest that the impact of the ionised wind is sufficient to inject enough energy to disturb the ordered gas motion within the host galaxy disc.

\subsection{Multi-phase outflow}\label{sec_discussion_multiphase_winds}
Exploiting the multi-band observations of NGC 424 we observed a clear emission line asymmetry in warm and intermediate-to-high ionization transitions such as the \NeIIImu, \OIII, and \NeVmu emission lines from recent MIRI observations and archival MUSE data. On the other hand, the line profile of the cold and warm molecular transitions (i.e., CO~(2-1) and \hiisi) considered in this work is always well-reproduced with a single Gaussian profile, suggesting that such transitions originate from ordered motions within the gaseous disc, with no evidence of molecular gas outflows. The detailed morphological and kinematical modeling of the warm ionised outflow presented in Sect. \ref{sec.3d_outflow_kin_model} allowed for the investigation of the redshifted part of the AGN outflow in NGC 424, leveraging the infrared wavelength to optical transitions. The outflow partially lies behind the galactic disc, a positioning that makes it more attenuated in the optical regime. 

We also investigated the highly-ionised counterpart of the outflow by fitting the spatially-integrated \NeVmu emission line (IP = 97 eV). Unfortunately, due to the low S/N of the broad component of this transition, we could not provide a detailed kinematical modeling as done for the \NeIIImu and \OIII transitions. Nevertheless, we estimate an outflow velocity of the highly-ionised phase traced by the \NeVmu emission consistent with the warm ionised components (see Sect. \ref{sec.highly_ionised_outflow} and Tab. \ref{tab:moka_outflow_fit}). 

Our findings on the the warm and highly ionised outflow rates as traced by the \OIII, \Halpha, \NeIIImu, and \NeVmu emission lines and presented in Sect. \ref{sec.outflow_energetic} suggest that the largest contribute to the mass outflow rate in NGC 424 has to be ascribed to the warm-ionised phase, as also confirmed by recent JWST MIRI observations of local AGN targeting the \NeVmu emission line \citep{Zhang2024}. Indeed, the mass outflow rate of the highly-ionised phase traced by the \NeVmu emission line is about an order of magnitude smaller (see Tab. \ref{tab:energetics}). Overall, our estimates for the mass outflow rate of the ionised phase are consistent with recent works studying the ionised outflow energetics in local Seyfert galaxies, showing mass outflow rates spanning the interval of $\rm \dot M_{out}$ $\sim$ 10$^{-2 \pm 0.8}$ \msun yr$^{-1}$ \citep[e.g.][]{Davies2020, Riffel2023, Zhang2024, Esposito2024}. Nevertheless, our estimate of the mass outflow rate in NGC 424 is $\sim$ two orders of magnitude smaller with respect to the value reported for the same target by \citet{Kakkad2022}. Indeed, studying the spatially resolved \OIII properties they found a mass outflow rate of $\rm \dot M_{out}$ = 26 $\pm$ 15 \msun yr$^{-1}$. On the other hand, they also provided mass outflow rate estimates computed from the integrated 3\arcsec fibre spectra and 1.5\arcsec $\times$ 10\arcsec integrated slit spectra, of 0.1 and 0.05 \msun yr$^{-1}$, respectively. We notice that our results are consistent with those presented in the integrated analysis of \citet{Kakkad2022} but not with the resolved estimates. The discrepancy between our resolved estimate and that presented in \citet{Kakkad2022} could be due to the amount of line flux associated to the outflow and its extension. Indeed, as shown in Fig. A4 in \citet{Kakkad2022}, they found outflowing gas up to extremely large scales compared to our analysis, therefore summing up fluxes from a larger number of pixels. This is also supported by the fact that both estimates are consistent when considering the values computed from the inner region of 3 $\arcsec$. Moreover, at variance with \citet{Kakkad2022}, we computed the \OIII ionised mass outflow rate only considering the emission extracted from the MIRI FOV, in order to provide consistent results with the Mid-IR analysis, and therefore neglecting part of the emission originating from the NE side of the nucleus. Finally, as presented in Sect. \ref{sec.outflow_energetic}, we also estimated the amount of ionised mass from \Halpha emission. This value is $\sim$ 4 times larger than the ionised mass traced by the \OIII emission line and still 1.5 dex smaller with respect to the value provided in \citet{Kakkad2022}. As discussed in \citet{Carniani2015}, the ionised mass traced by the \OIII emission line has to be considered as a lower limit of the total ionised mass when compared to the mass traced by recombination lines as \Halpha or \Hbeta \citep{Karouzos2016, Rakshit2018}.

\section{Conclusions}\label{sec.conclusion}
In this work we presented the first observations of the Mid-InfraRed Activity of Cicumnuclear Line Emission (MIRACLE) program, aimed at tracing the Mid-IR radiation exploiting MIRI/MRS observations of the circumnuclear region of local AGN in the 5-28 $\mu$m wavelength range. We completed the multi-phase analysis of the NGC 424 galaxy aided by archival MUSE and ALMA data, tracing the warm ionised and cold molecular phases. The main conclusions from this work are the following:
\begin{itemize}
    \item We traced the ionised (\NeIIImu, \OIII), warm (\hiisi) and cold (CO~(2-1)) molecular gas phases in the galaxy disc revealing the intrinsic rotation curve from a few pc up to $\sim$ 5 kpc scales and providing an accurate estimate of the dynamical mass of the NGC 424 galaxy of M$_{\rm dyn}$ = 1.09 $\pm$ 0.08 $\times$ 10$^{10}$ \msun.
    \item We reported the first evidence of the co-existence of an elongated velocity dispersion structure ($\sigma$ $\geq$ 350 \kms) perpendicular to the direction of the ionisation-cone ionised outflow in both the ionised, cold and warm molecular phases. At variance with previous works, we mainly ascribe such scenario to a tight interplay between the gaseous disc and the outflow, possibly due to the impact of wind on the host, thus causing highly turbulent gas in the perpendicular direction, without necessarily having a low-inclination radio-jet.
    \item We spatially resolved and modeled the ionised gas phase in the outflow traced by the broad component of the \NeIIImu and \OIII emission lines, from MIRI and MUSE IFS data, respectively. We estimated outflow velocities up to 10$^3$ \kms from both tracers and mass outflow rates of the order of 10$^{-1.0 \pm 0.4}$ \msun yr$^{-1}$. Moreover, the 3D morphological and kinematical modeling support the scenario of a tight interaction between the ionised outflow and the gaseous galactic disc, suggesting that the impact of the outflow on the disc might be as relevant as the impact of the jet in producing the observed enhancement of the velocity dispersion perpendicular to wind direction.
    \item We fit the spatially integrated \NeVmu emission line tracing the highly ionised outflow in NGC 424 and found an average wind velocity consistent with the warm ionised phase and a mass outflow rate of the highly-ionised phase of 1.6 $\times$ 10$^{-2}$ \msun yr$^{-1}$.
\end{itemize}
In conclusion, in this work we exploited MRS data and archival MUSE and ALMA data to shed light on the physical properties of the NGC 424 active galaxy across various phases and scales. In particular, we carried out a detailed analysis of the ionised outflow properties traced by Mid-IR and optical emission lines, evaluating the energetic impact of the wind onto the host galaxy. Finally, with this work we remarked the necessity of carrying out a multi-wavelength analysis of AGN outflows, aided by optical, millimetre, and Mid-IR observations to unveil the total energetics of outflows. 
\begin{acknowledgements}
CM, GC, AM, FM, FB, EB, GV and AF acknowledge the support of the INAF Large Grant 2022 "The metal circle: a new sharp view of the baryon cycle up to Cosmic Dawn with the latest generation IFU facilities". CM, GC, AM, GT, FM, FB, EB also acknowledge the support of the grant PRIN-MUR 2020ACSP5K\_002 financed by European Union – Next Generation EU. EDT is supported by the European Research Council (ERC) under grant agreement no. 101040751. AM, FM, GC, IL acknowledge support from project PRIN-MUR project “PROMETEUS”  financed by the European Union -  Next Generation EU, Mission 4 Component 1 CUP B53D23004750006. GV and SC acknowledge funding from the European Union (ERC, WINGS,101040227). G.T. acknowledges financial support from the European Research Council (ERC) Advanced Grant under the European Union’s Horizon Europe research and innovation programme (grant agreement AdG GALPHYS, No. 101055023). GS acknowledges the INAF Mini-grant 2023 TRIESTE (“TRacing the chemIcal hEritage of our originS: from proTostars to planEts”; PI: G. Sabatini), the project ASI-Astrobiologia 2023 MIGLIORA (Modeling Chemical Complexity, F83C23000800005), the National Recovery and Resilience Plan (NRRP), Mission 4, Component 2, Investment 1.1, Call for tender No. 104 published on 2.2.2022 by the Italian MUR, funded by the European Union – NextGenerationEU – Project 2022JC2Y93 ``Chemical Origins: linking the fossil composition of the Solar System with the chemistry of protoplanetary disks'' (CUP J53D23001600006 - Grant Assignment Decree No. 962 adopted on 30.06.2023 by the Italian MUR), and the PRIN-MUR 2020 BEYOND-2p (Astrochemistry beyond the Second period elements, Prot. 2020AFB3FX). MM is thankful for support from the European Space Agency (ESA). AVG acknowledges support from the Spanish grant PID2022-138560NB-I00, funded by MCIN/AEI/10.13039/501100011033/FEDER, EU. FS acknowledges financial support from the PRIN MUR 2022 2022TKPB2P - BIG-z, Ricerca Fondamentale INAF 2023 Data Analysis grant ``ARCHIE ARchive Cosmic HI \& ISM  Evolution'', Ricerca Fondamentale INAF 2024 under project "ECHOS" MINI-GRANTS RSN1. MVZ acknowledges partial financial support from the "Fondazione CR Firenze" with the program "Ricercatori a Firenze 2023”. This paper makes use of the following ALMA data: ADS/JAO.ALMA 2021.1.01150.S. ALMA is a partnership of ESO (representing its member states), NSF (USA) and NINS (Japan), together with NRC (Canada), NSTC and ASIAA (Taiwan), and KASI (Republic of Korea), in cooperation with the Republic of Chile. The Joint ALMA Observatory is operated by ESO, AUI/NRAO and NAOJ.
\end{acknowledgements}

\bibliographystyle{aa}
\bibliography{example} 

\begin{appendix} 
\section{MIRI/MRS PSF modeling}\label{sec_app_PSF_miri_sub}
\begin{figure}
\centering
	\includegraphics[scale=0.6]{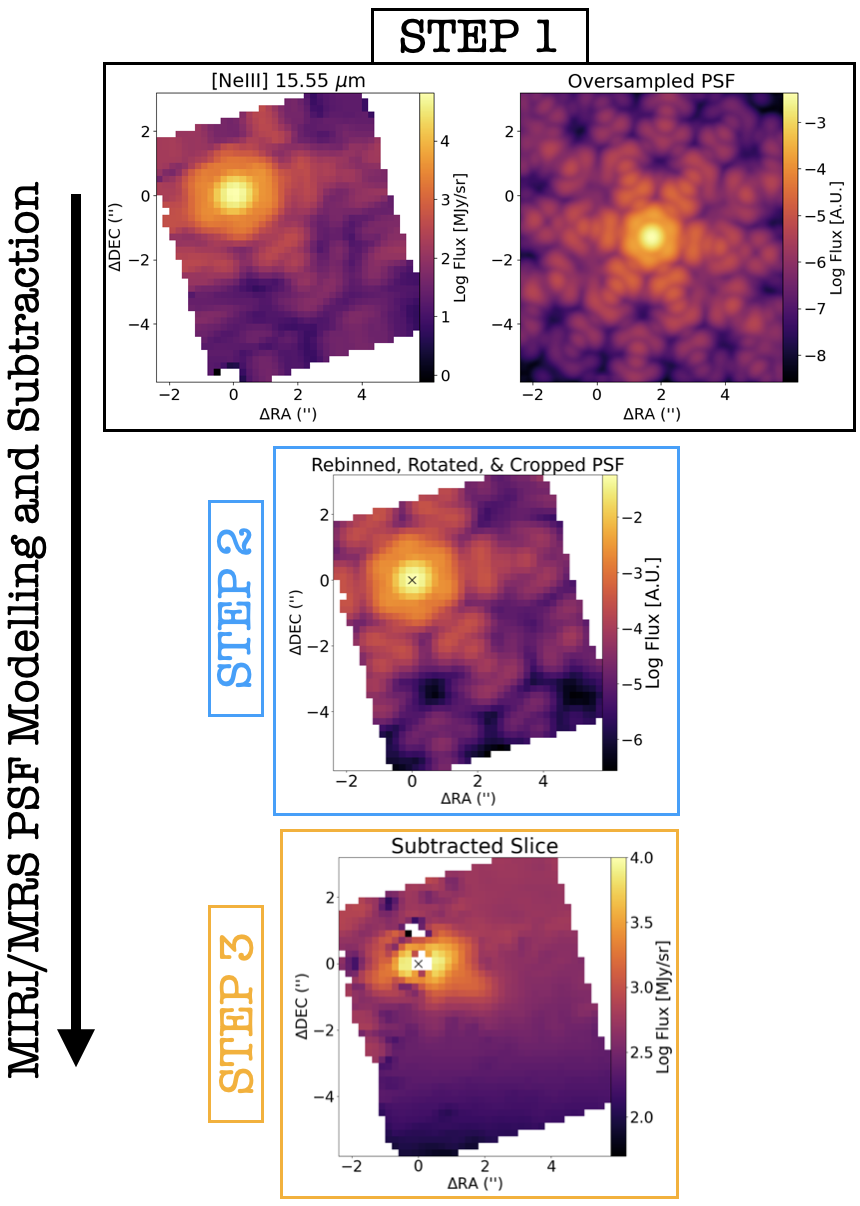}
    \caption{Scheme of the MIRI/MRS PSF modeling and subtraction applied to NGC 424 data. The arrow indicate the direction we followed during the PSF subtraction procedure. A detailed description of the various steps is presented in App. \ref{sec_app_PSF_miri_sub}. From left to right: A slice of MIRI Ch3 LONG data at the wavelength of the peak of the \NeIIImu emission line with the corresponding slice of the oversampled \textit{WebbPSF} cube. Step 2 and 3 represent the centred, rebinned, rotated and cropped PSF and the final PSF subtracted data slice.}
    \label{fig:psf_model_scheme}
\end{figure}

Figure \ref{fig:psf_model_scheme} shows a scheme of the PSF wavelength-dependent subtraction procedure that we performed to model and subtract the PSF to each slice of the pipeline product data cubes. Indeed, as shown by Eq.~1 and Fig.~2 in \citet{Law2023_drizzle}, the average FWHM of the MIRI/MRS PSF varies across the four channels spanning the range FWHM = 0.4\arcsec, 0.5\arcsec, 0.6\arcsec, 0.9\arcsec, in Ch1, Ch2, Ch3, Ch4, respectively. In this section we discuss our procedure applied to the MIRI/MRS Ch3 LONG data, the same applies with the proper wavelength-dependent changes to all the other sub-bands data-cubes. First, we downloaded the MIRI MRS PSF model cube using the \textit{WebbPSF} tool v.1.4.0 \citep{Perrin2014}, which provide a model cube spatially oversampled of a factor of 4 with respect to the observations (i.e. 0.05 \arcsec/pixel for Ch3). We performed a wavelength-dependent alignment, rotation and flux normalization between the PSF and our data cube, following three main steps. In particular, we used the oversampled PSF cube to align the PSF and data centroid at sub-pixel scale performing a 2D Gaussian fit of the data and the PSF at each slice. We then aligned the PSF by centering it on the data centroid and rotating it of the proper rotation angle between the sky coordinates and the IFU slicer-plane. Then, we binned the PSF 2D image to the data spatial scale (e.g. 0.2 \arcsec/pxl for Ch3 sub-bands) and cropped the edges to match the data FOV and avoid artifacts at the outskirts of the data image. To account for the flux scaling factor between the PSF image and the corresponding data slice, we performed a 2D Gaussian fit to the PSF slice with four free parameters, i.e. the centroid (X$_{PSF}$, Y$_{PSF}$) and its FWHM ($\sigma_X$, $\sigma_Y$). Subsequently, we repeated the same procedure for the corresponding data slice, using as fixed values the $\sigma_X$ and $\sigma_Y$ inferred from the PSF fit and using an additional 2D offset surface applied to the data to account for the background emission. We measured the amplitude of the 2D gaussian fitted to the PSF and data slices and used it to normalize the PSF intensity in each spaxel. Finally, we subtracted to each data slice the obtained PSF. The final result is a PSF-subtracted data cube which allows us to investigate the spatially resolved properties of gas emission across the FOV.

As stressed at the beginning of this section, due to the wavelength-dependent PSF FWHM \citep{Law2023_drizzle, Argyriou2023, Bajaj2024} we used a tailored PSF model cube to subtract the PSF emission from all the 12 MIRI/MRS data-cubes, repeating the detailed modeling described above to each slice of each data-cube.




\section{Electron density and dust attenuation maps}\label{app_sec_electron_density}
In this appendix we present the electron density estimate for the warm ionised gas obtained from the total intensities of the \SIIall emission lines and the dust extinction map for NGC 424 making use of the Balmer decrement \Halpha/\Hbeta, assuming a \citet{Calzetti2000} attenuation law. In particular, to compute the electron density we used the \SII[6717]/\SII[6731] diagnostic line ratio \citep{Osterbrock2006} from spaxels with S/N $\geq$ 5 on the total \SIIall emission lines, and assuming a typical value for the temperature of ionised gas of T$_{\rm e}$ = 10$^4$ K.

\begin{figure*}
    \centering
    \includegraphics[scale=0.7]{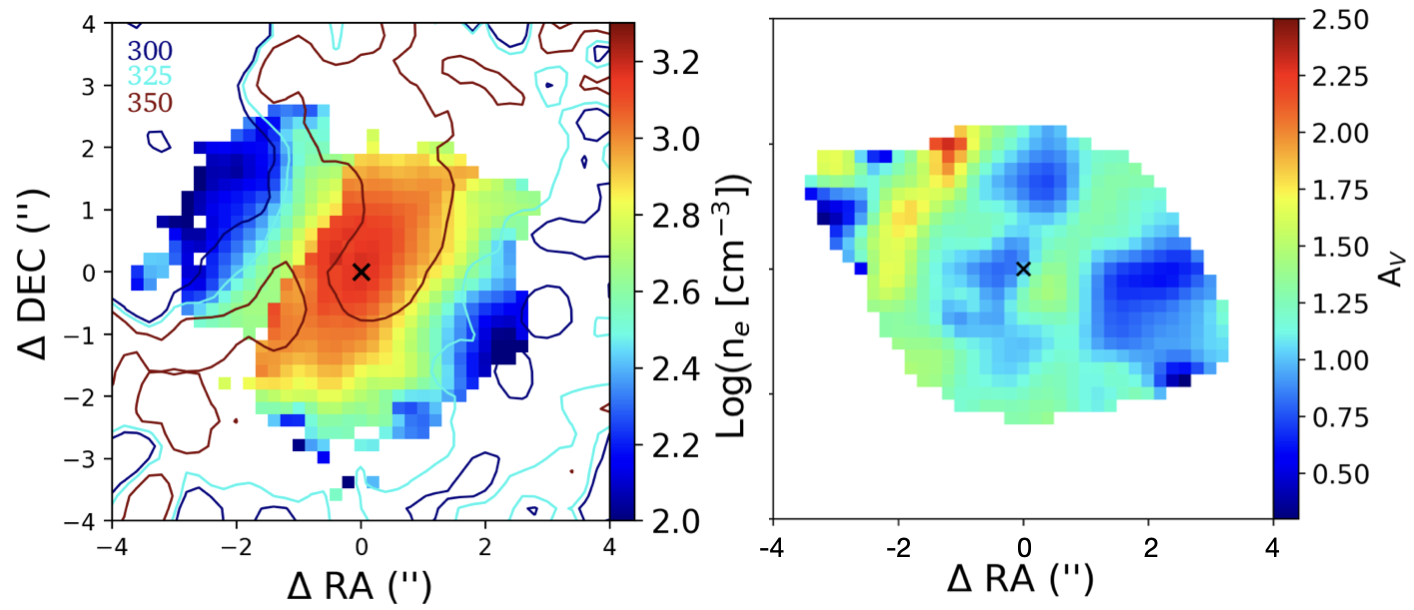}
    \caption{Spatially resolved estimate of the ionised gas electron density and dust attenuation in the circumnuclear region of NGC 424 obtained from the \SIIall doublet and \Halpha/\Hbeta ratio of MUSE data. Spaxels at S/N $\leq$ 5 are masked. Left panel: Red, cyan, and blue contours represent the 350, 325, and 300 km s$^{-1}$ contours of the broad component of the \OIII velocity dispersion map shown in Fig. \ref{fig:outflow_moka_OIII}.}
    \label{fig:eletron_density_muse}
\end{figure*}

\section{Multi-phase kinematical maps of the \MOKA \ fit}\label{sec_app_moka_fit}
The best fit models obtained fitting with tailored \MOKA \ modeling the multi-phase gas disc emission, as traced by \hiisi, \NeIIImu, CO~(2-1), and \OIII emission lines, observed with MIRI, ALMA, and MUSE, respectively, are shown in this section. Details of the modeling are presented in Sect. \ref{sec.3d_disc_kin_model} and the best-fit parameters are listed in Tab. \ref{tab:moka_disc_fit}. For more details on the \MOKA \ modeling routine see \citet{Marconcini2023} and Marconcini et. al 2025 (in press).

\begin{figure*}
\centering
\includegraphics[scale=0.4]{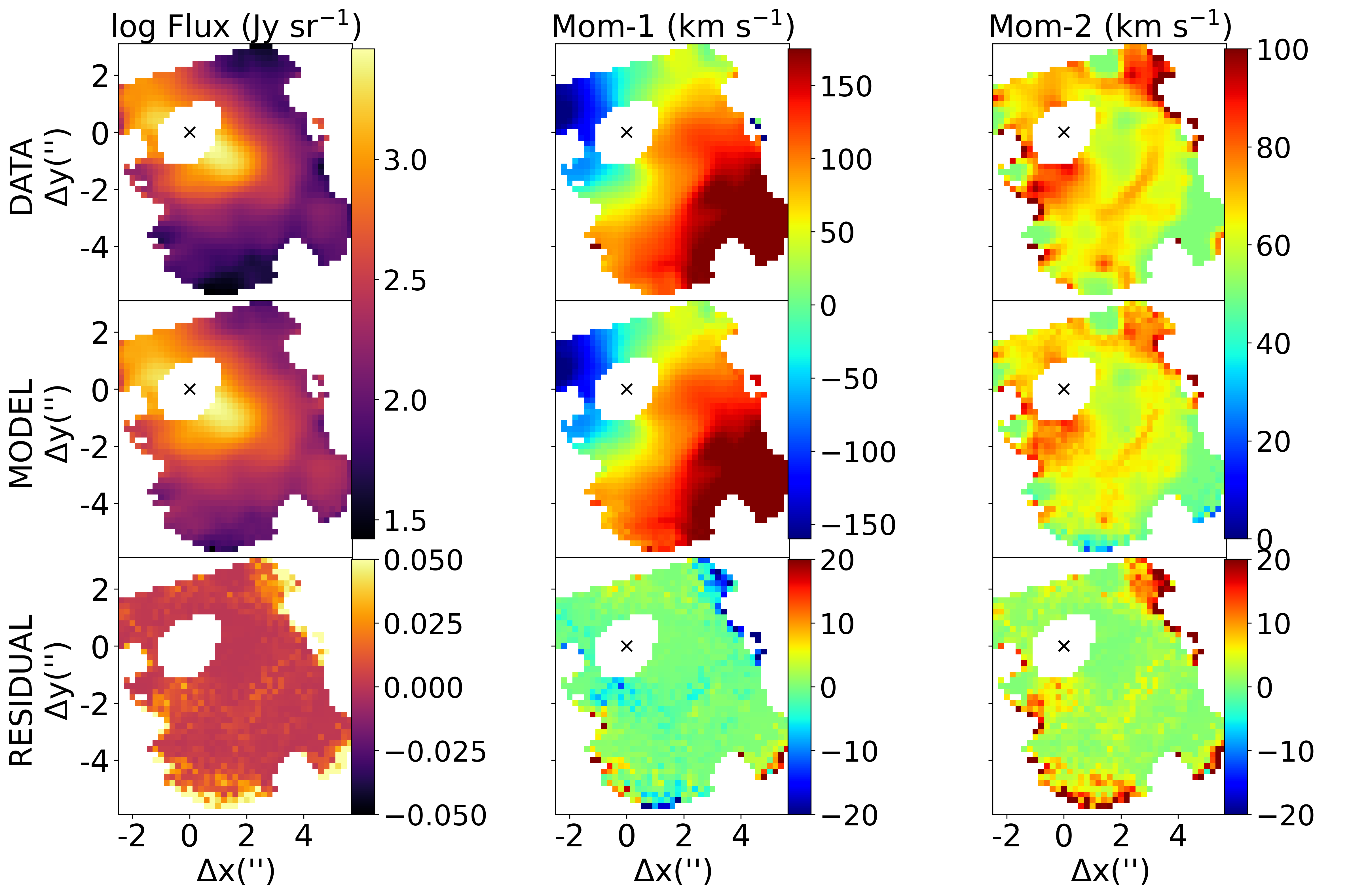}
    \caption{Comparison of the observed (top panel) and best-fit \MOKA \ (middle panel) moment maps for the \hiisi emission line. Residuals are shown in the bottom panel. The modeling and best-fit parameters are presented in Sect. \ref{sec.3d_disc_kin_model} and listed in Tab. \ref{tab:moka_disc_fit}.}
    \label{fig:MOKA_H2_momamps}
\end{figure*}

\begin{figure*}
\centering
\includegraphics[scale=0.4]{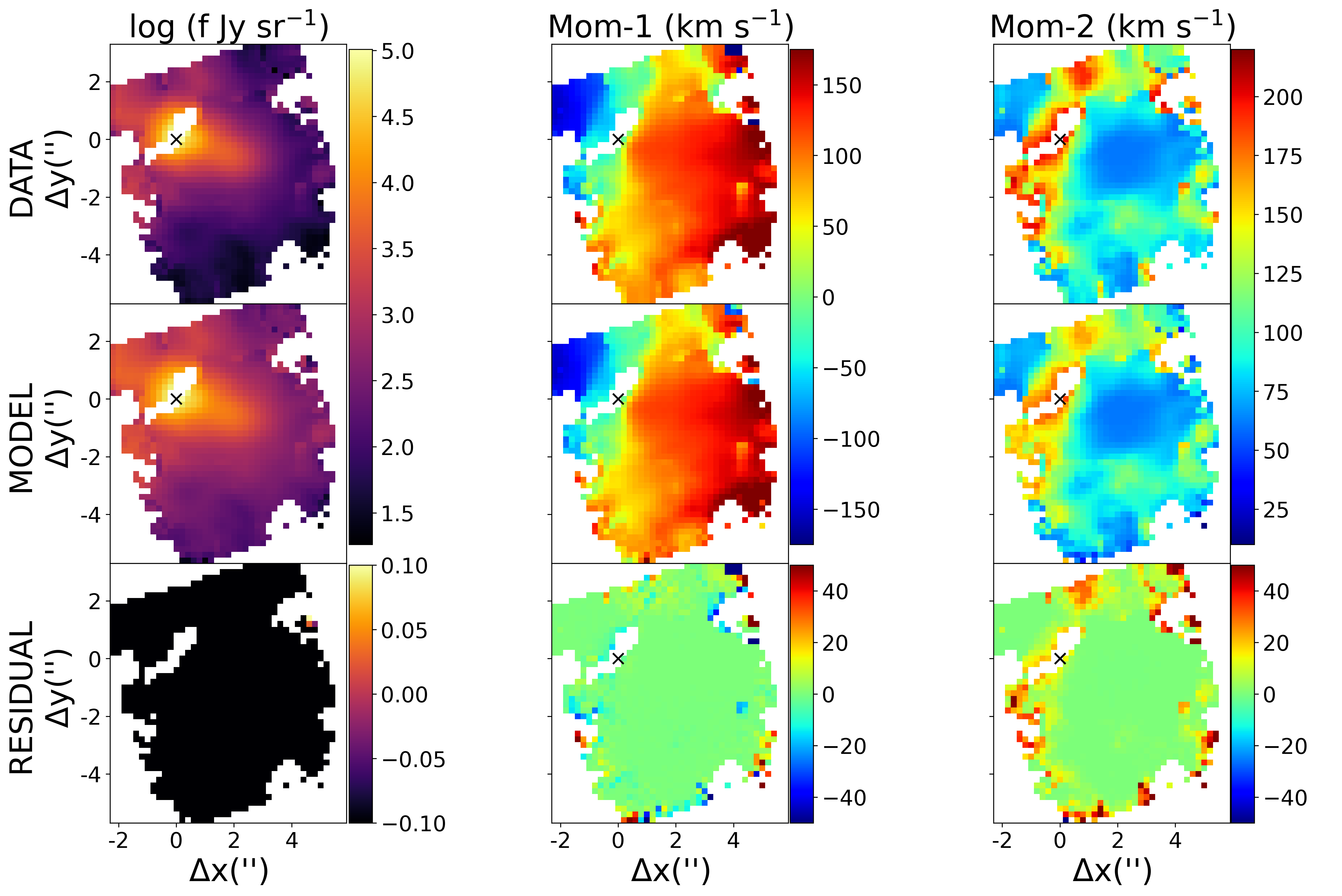}
    \caption{Continue of Fig. \ref{fig:MOKA_H2_momamps} for the \NeIIImu emission line.}
    \label{fig:MOKA_NeIII_momamps}
\end{figure*}

\begin{figure*}
\centering
\includegraphics[scale=0.4]{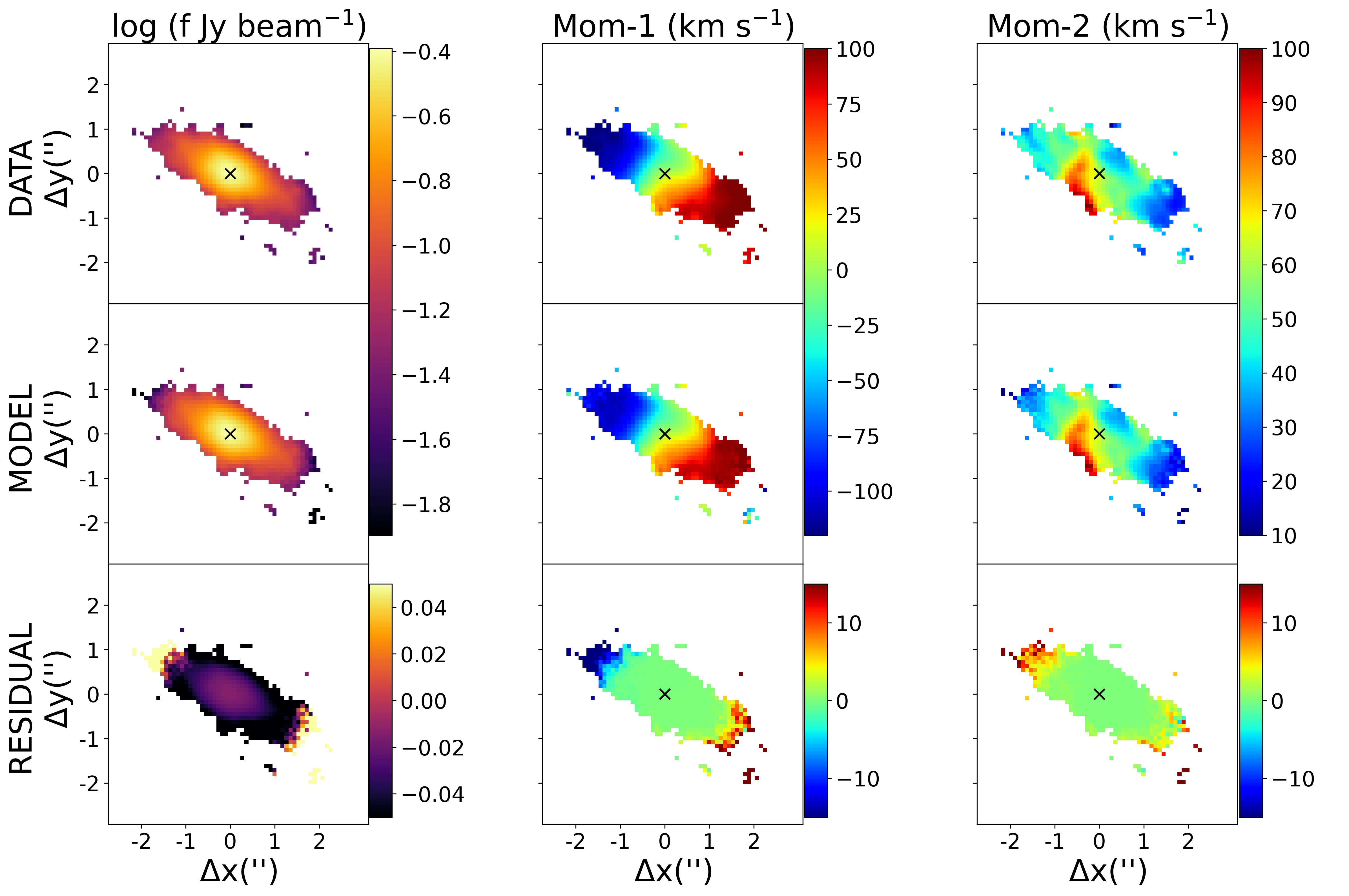}
    \caption{Continue of Fig. \ref{fig:MOKA_H2_momamps} for the CO~(2-1) emission line observed with ALMA.}
    \label{fig:MOKA_CO21_momamps}
\end{figure*}

\begin{figure*}
\centering
\includegraphics[scale=0.4]{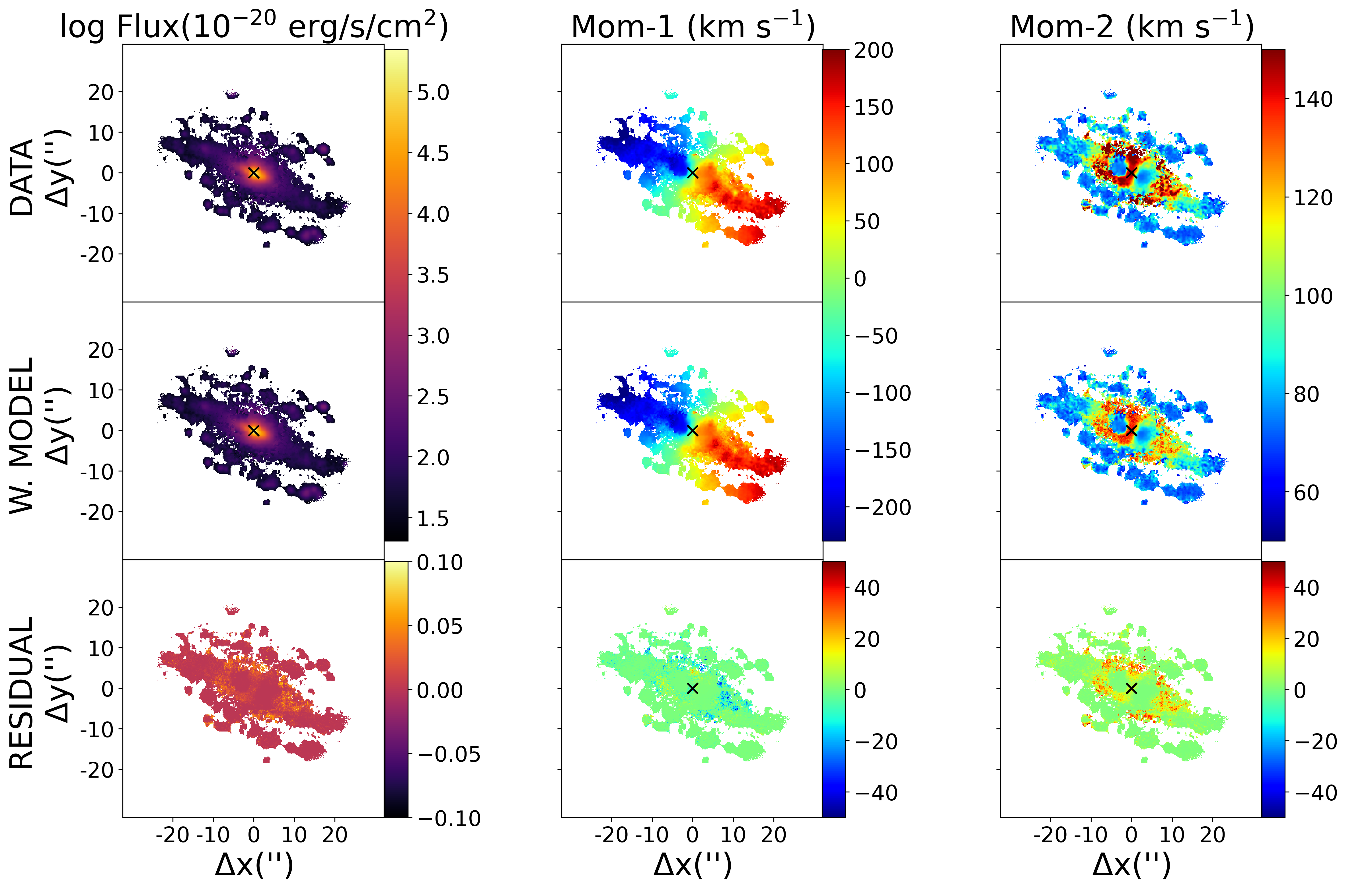}
    \caption{Continue of Fig. \ref{fig:MOKA_H2_momamps} for the narrow component of the \OIII emission line observed with MUSE.}
    \label{fig:MOKA_OIII_momamps}
\end{figure*}

\end{appendix}

\end{document}

%% file: fde_emlines.tex
\usepackage{xargs}
\usepackage{xspace}
\usepackage{textgreek}

\newcommand{\Halpha}{\text{H\textalpha}\xspace}
\newcommand{\Hbeta}{\text{H\textbeta}\xspace}

\newcommandx{\permittedEL}[6][1=O,2=III,3=,4=,5=,6=]{\text{{#1}\,{\sc {#2}}{#3}{#4}{#5}{#6}}\xspace}
\newcommandx{\semiforbiddenEL}[6][1=O,2=III,3=,4=,5=,6=]{\text{{#1}\,{\sc{#2}}]{#3}{#4}{#5}{#6}}\xspace}
\newcommandx{\forbiddenEL}[6][1=O,2=III,3=,4=,5=,6=]{\text{[{#1}\,{\sc{#2}}]{#3}{#4}{#5}{#6}}\xspace}

\newcommandx{\NVL}[1][1=1243]{\permittedEL[N][v][\textlambda][#1]}
\newcommandx{\NVall}{\permittedEL[N][v][\textlambda][\textlambda][1239,][1243]}

\newcommandx{\CIIL}[1][1=232x]{\semiforbiddenEL[C][ii][\textlambda][#1]}
\newcommandx{\CIIall}{\semiforbiddenEL[C][ii][\textlambda][\textlambda][2324--][2329]}

\newcommandx{\NIVL}[1][1=1486]{\semiforbiddenEL[N][iv][\textlambda][#1]}

\newcommandx{\CIVL}[1][1=1550]{\permittedEL[C][iv][\textlambda][#1]}

\newcommandx{\HeIIL}[1][1=1640]{\permittedEL[He][ii][\textlambda][#1]}

\newcommandx{\semiOIIIL}[1][1=1666]{\semiforbiddenEL[O][iii][\textlambda][#1]}

\newcommandx{\NIIIL}[1][1=1750]{\semiforbiddenEL[N][iii][\textlambda][#1]}

\newcommandx{\CIII}{\semiforbiddenEL[C][iii]}
\newcommandx{\CIIIL}[1][1=1909]{\semiforbiddenEL[C][iii][\textlambda][#1]}

\newcommandx{\NeIVL}[1][1=2424]{\forbiddenEL[Ne][iv][\textlambda][#1]}

\newcommandx{\MgIIL}[1][1=2803]{\permittedEL[Mg][ii][\textlambda][#1]}

\newcommandx{\NeVL}[1][1=3426]{\forbiddenEL[Ne][v][\textlambda][#1]}

\newcommandx{\OIIL}[1][1=3727]{\forbiddenEL[O][ii][\textlambda][#1]}

\newcommand{\NeIII}{\forbiddenEL[Ne][iii]}
\newcommandx{\NeIIIL}[1][1=3869]{\forbiddenEL[Ne][iii][\textlambda][#1]}

\newcommandx{\NeIIImu}[0][]{\forbiddenEL[Ne][iii][][][]}
\newcommandx{\NeVmu}[0][]{\forbiddenEL[Ne][v][][][]}

\newcommandx{\OIVmu}[1][1=25.8]{\forbiddenEL[O][iv][][#1][\textmu m]}

\newcommand{\OIII}{\forbiddenEL[O][iii]}
\newcommandx{\OIIIL}[1][1=5007]{\forbiddenEL[O][iii][\textlambda][#1]}
\newcommand{\OIIIall}{\forbiddenEL[O][iii][\textlambda][\textlambda][4959,][5007]}

\newcommandx{\NIL}[1][1=5200]{\forbiddenEL[N][i][\textlambda][#1]}

\newcommandx{\OIL}[1][1=6300]{\forbiddenEL[O][i][\textlambda][#1]}

\newcommandx{\HeIL}[1][1=5875]{\permittedEL[He][i][\textlambda][#1]}

\newcommandx{\NIIL}[1][1=6584]{\forbiddenEL[N][ii][\textlambda][#1]}
\newcommand{\NIIall}{\forbiddenEL[N][ii][\textlambda][\textlambda][6549,][6584]}

\newcommand{\SII}{\forbiddenEL[S][ii]}

\newcommand{\SIIall}{\forbiddenEL[S][ii][\textlambda][\textlambda][6716,][6731]}

\newcommandx{\OIIAuL}[1][1=7325]{\forbiddenEL[O][ii][\textlambda][#1]}

\newcommandx{\CIIFIRL}{\forbiddenEL[C][ii][\textlambda][158\,\mum]}


%% file: NGC424_commissioning.tex
\begin{table*}
    \begin{center}
    \caption{Summary of MIRI/MRS observations for NGC 424 as part of the MIRACLE program.
    }\label{tab:MIRI_commissioning_NGC424}
    \setlength{\tabcolsep}{5pt}
    \begin{tabular}{cccccccc}
  \hline
   Target & MRS band & Detector &  Groups & Integrations & Exposures & Dithers & Exposure Time (s) \\
    (1) & (2) &  (3) & (4) & (5) & (6) & (7) & (8) \\
   \hline
   NGC 424 & SHORT, MEDIUM, LONG & SHORT, LONG &  25 & 8 & 1 & 4 & 2297.7\\
   Background & SHORT, MEDIUM, LONG & SHORT, LONG &  25 & 8 & 1 & 4 & 2297.7\\

  \hline
  \end{tabular}
  \end{center}
Different columns describe the pointing target of the MRS (1), the multi-band configuration used (2), the detector (3), the number of groups (4), integrations (5), exposures (6), dithers (7) for each observation of the target. The total exposure time (8) is referred to each band and detector configuration. See program ID 6138 for more details.
\end{table*}

%% file: line_fluxes.tex
\begin{table*}[t]
    \begin{center}
    \caption{Mid-IR emission line fluxes from MIRI/MRS channels, in units of MJy/sr.
    }\label{tab:line_fluxes_miri}
    \setlength{\tabcolsep}{4pt}
    \begin{tabular}{llcp{1.5cm}llcllc}
  \hline
\hline

   & Line & Wave &  Channel & I.P. &  N$_{comps}$ & Flux & Uncertainty \\
    \hline
   &   & [$\mu$m] &  & [eV] & & [MJy/sr] & [MJy/sr] \\
   \hline
   \multirow{5}{*}
    & \hiisi & 17.0350 & 3C & ---- & 1  & 1.4 $\times$ 10$^{7}$   & 3.2 $\times$ 10$^{5}$\\
    & \hiisii & 12.2790 & 3A & ---- & 1  & 1.9 $\times$ 10$^{6}$ & 5.7 $\times$ 10$^{4}$\\
    & \hiisiii & 9.6649 & 2B & ---- & 1  & 2.4 $\times$ 10$^{6}$& 7.6 $\times$ 10$^{4}$\\
    & \hiisiiii & 8.0258 & 2A & ---- & 1  & 4.6 $\times$ 10$^{5}$  & 9.7 $\times$ 10$^{3}$\\
    & \hiisiiiii & 6.9091 & 1C & ---- & 1  & 1.1 $\times$ 10$^{6}$  & 1.9 $\times$ 10$^{4}$\\
    & [SIV] & 10.5105 & 2C & 34.79 & 2  & 2.2 $\times$ 10$^{7}$  & 3.5 $\times$ 10$^{5}$\\
    & [SIII] & 18.7130 & 4A & 23.34  & 2  & 7.7 $\times$ 10$^{6}$  & 2.4 $\times$ 10$^{5}$\\

   & [ArII] & 6.9853 & 1C & 15.76 & 2  & 4.9 $\times$ 10$^{6}$  & 7.1 $\times$ 10$^{7}$\\
   & [ArIII] & 8.9914 & 2B & 27.63 & 2  & 4.1 $\times$ 10$^{6}$  & 6.5 $\times$ 10$^{4}$\\
    & [ArV] & 13.1022 & 2B & 59.81 & 1  & 5.0 $\times$ 10$^{5}$  & 7.9 $\times$ 10$^{3}$\\
   & [MgV] & 5.6099 & 1A & 109.24 & 2  & 2.6 $\times$ 10$^{6}$  & 3.7 $\times$ 10$^{4}$\\
   & [MgVII] & 5.5032 & 1A & 186.51 & 2  & 1.3 $\times$ 10$^{6}$  & 1.9 $\times$ 10$^{4}$\\

   & [NeVI] & 7.6524 & 2A &  126.21 & 2  & 8.6 $\times$ 10$^{6}$  & 1.3 $\times$ 10$^{5}$\\

   & [NeII] & 12.8136 & 3A &  21.56 & 2  & 2.3 $\times$ 10$^{7}$  & 3.9 $\times$ 10$^{5}$\\
   & [NeV] & 14.3217 & 3B & 97.12  & 2  & 4.2 $\times$ 10$^{7}$  & 5.8 $\times$ 10$^{5}$\\
   & [NeIII] & 15.5550 & 3C & 40.96  & 2  & 6.4 $\times$ 10$^{7}$  & 9.7 $\times$ 10$^{5}$\\
   & [NeV] & 24.3175 & 4B & 97.12 & 2  & 1.4 $\times$ 10$^{7}$  & 3.8 $\times$ 10$^{5}$\\
   & [OIV] & 25.8903 & 4C & 54.93  & 2  & 7.4 $\times$ 10$^{7}$  & 1.9 $\times$ 10$^{6}$\\

   & FeII & 5.3402 & 1A & 7.90  & 2  & 1.5 $\times$ 10$^{6}$  & 2.8 $\times$ 10$^{4}$\\
    & FeVIII & 5.4466 & 1A & 124.80  & 2  & 2.4 $\times$ 10$^{7}$  & 2.8 $\times$ 10$^{5}$\\
   & FeVII & 9.5267 & 2B & 99.10  & 2  & 1.5 $\times$ 10$^{6}$  & 1.7 $\times$ 10$^{4}$\\

  \hline
   \hline

  \end{tabular}
  \end{center}
From left to right: Line name, rest-frame wavelength, MIRI/MRS sub-channel, ionisation potential of the transition, maximum number of Gaussian components used to fit the line profile. Fluxes are obtained integrating the best-fit multi-Gaussian model of the reduced data-cubes, without subtracting the PSF, and considering spaxels at S/N $\geq$ 2.
\end{table*}

%% file: mokafit_disc.tex
\begin{table}[t]
    \begin{center}
    \caption{Best-fit parameters for the disc properties in NGC 424, inferred with the \MOKA\ framework (see also Sec.~\ref{sec.3d_disc_kin_model}).
    }\label{tab:moka_disc_fit}
    \setlength{\tabcolsep}{3pt}
    \begin{tabular}{llcp{1.5cm}llcllc}
  \hline
   & Parameter & Free &  \hiisi & \NeIIImu & CO(2-1) & \OIII \\
   \hline
   \multirow{5}{*}
   & $\beta$ & Y & 55 $\pm$ 4 & 65 $\pm$ 9 & 58 $\pm$ 4 & 60 $\pm$ 9 \\
   & V$_{rot}$ & Y & 188 $\pm$ 38 & 176 $\pm$ 44 & 90 $\pm$ 42  & 198 $\pm$ 46\\
   & V$_{disp}$ & Y & 60 $\pm$ 25 & 75 $\pm$ 20 & 75 $\pm$ 15  & 65 $\pm$ 10\\
   \hline
   & P.A. & N & 28  & 28  & 28  & 28 \\
   & N$_{shell}$ & N & 9  & 9  & 8  & 20 \\

  \hline
  \end{tabular}
  \end{center}
From left to right: Parameter name, label to specify if the parameter is free (Y) or fixed (N), and tracer of a specific gas phases fitted with \MOKA. The optimized parameters are the disc inclination ($\beta$), the circular velocity (V$_{rot}$), and the disc intrinsic velocity dispersion (V$_{disp}$). The disc position angle (P.A.) is kept fixed during the fit. For a comprehensive discussion of each model parameter see \citet{Marconcini2023}. N$_{shell}$ is the number of concentric circular shells used to reproduce the observed features. Here we list the disc inclination averaged across the radius and the maximum disc circular velocity inferred with \MOKA. A detailed discussion of the disc rotation curve is presented in Sec. \ref{sec.3d_disc_kin_model} and shown in Fig. \ref{fig:disc_v_profile}.
\end{table}

%% file: mokafit_outflow.tex
\begin{table}[t]
    \begin{center}
    \caption{Best-fit parameters for the ionised outflow properties in NGC 424, traced by \NeIIImu and \OIIIL emission lines and inferred with the \MOKA\ framework.
    }\label{tab:moka_outflow_fit}
    \setlength{\tabcolsep}{4pt}
    \begin{tabular}{llcp{2.5cm}llcllc}
  \hline
   & Parameter & Free & \NeIIImu & \OIIIL \\
   \hline
   \multirow{5}{*}
   & Inclination ($\beta$) & Y &  25 $\pm$ 7 & 35 $\pm$ 3   \\
   & V$_{out}$ & Y & 975 $\pm$ 30  & 990 $\pm$ 35\\
   & V$_{disp}$ & Y &  90 $\pm$ 10  & 95 $\pm$ 25\\
   \hline
   & P.A. ($\gamma$) & N &  118  & 118 \\
   & R$_{max}$ & N &  1.1  & 1.6 \\
  \hline
  \end{tabular}
  \end{center}
From left to right: Parameter name, label to specify if the parameter is free (Y) or fixed (N), and tracer of a specific gas phases fitted with \MOKA. The optimised parameters are the outflow inclination ($\beta$), the outflow radial velocity (V$_{out}$), and the global conical outflow intrinsic velocity dispersion (V$_{disp}$). The outflow inclination values are referred as the inclination of the cone axis with respect to the plane of the sky.  The outflow position angle ($\gamma$) is kept fixed during the fit. R$_{max}$ is the maximum outflow radius in kpc. For details on the conical outflow structure and parameters see Sec.~\ref{sec.3d_outflow_kin_model}.
\end{table}

%% file: energetics.tex
\begin{table*}
    \begin{center}
    \caption{Summary of outflow energetic properties in NGC 424.
    }\label{tab:energetics}
    \setlength{\tabcolsep}{5pt}
    \begin{tabular}{ccccccc}
  \hline
  \vspace{1mm}
   Line ID &  M$_{\rm out}$ &  $\rm \dot M_{out}$ & E$_{\rm kin}$ & L$_{\rm kin}$ & $\rm \dot p_{out}$ \\
    -- &  10$^{4}$ \msun & 10$^{-2}$ \msun yr$^{-1}$ & 10$^{53}$ erg & 10$^{39}$ erg s$^{-1}$ & 10$^{32} dyne$ \\
   \hline
   \NeVmu  & 1.9 $\pm$ 0.9 &  1.6 $\pm$ 0.8 & 1.5 $\pm$ 0.7 & 4.1 $\pm$ 1.9 & 0.9 $\pm$ 0.4 \\
   \NeIIImu  & 5.4 $\pm$ 2.7 &  4.5 $\pm$ 2.3 & 5.1 $\pm$ 2.6 & 13 $\pm$ 7 & 2.7 $\pm$ 1.4 \\
   \OIIIL  & 24 $\pm$ 11  & 22 $\pm$ 10 & 23 $\pm$ 11 & 68 $\pm$ 31 & 14 $\pm$ 6 \\

  \hline
  \end{tabular}
  \end{center}
  NGC 424 outflow energetic properties traced by optical (\OIIIL) and Mid-IR (\NeVmu, \NeIIImu) emission lines. From left to right, columns are defined as follows: outflowing gas mass (M$_{\rm out}$), mass outflow rate ($\rm \dot M_{out}$), outflow kinetic energy (E$_{\rm kin}$), kinetic luminosity (L$_{\rm kin}$), and momentum rate ($\rm \dot p_{out}$). Quantities are computed using Eqs. \ref{eq.m_out}, \ref{eq.m_out_rate}, \ref{eq.outflow_mass_neon}.
\end{table*}